\documentclass[11pt,a4paper]{article}
\usepackage{jheppub}
\usepackage{graphicx}
\usepackage{slashed}
\usepackage{braket}
\usepackage{dsfont}
\usepackage{appendix}
\usepackage{multirow}
\usepackage{comment}
\usepackage{adjustbox}
\usepackage{cleveref}
\usepackage{xspace}
\usepackage{color}
\usepackage{float}
\usepackage[normalem]{ulem}

\newcommand{\LamQCD}{\Lambda_{\rm QCD}}
\newcommand{\LamNP}{\Lambda_{\rm NP}}
\newcommand{\as}{\alpha_s}
\newcommand{\cVR}{\cos(\delta_{R})}
\newcommand{\cS}{\cos(\delta_{ST})}
\newcommand{\cP}{\cos(\delta_{PT})}
\newcommand{\nub}{{\bar{\nu}}}

\newcommand{\Vcb}[0]{\ensuremath{|V_{cb}|}\xspace}

\makeatletter
\g@addto@macro\bfseries{\boldmath}
\let\Hy@backout\@gobble
\makeatother

\title{New Physics in inclusive semileptonic $B$ decays}

\author[a]{Alexandre Carvunis,} 
\author[a]{Gael Finauri,}
\author[a]{Paolo Gambino,}
\author[a]{Martin Jung,}
\author[a]{Sandro Mächler}
\affiliation[a]{Dipartimento di Fisica, Universit\`a di Torino \& INFN, Sezione di Torino, \\
Via Pietro Giuria 1, I-10125 Turin, Italy}

\emailAdd{alexandre.carvunis@unito.it}
\emailAdd{gael.finauri@unito.it}
\emailAdd{paolo.gambino@unito.it}
\emailAdd{martin.jung@unito.it}
\emailAdd{sandro.maechler@gmail.com}

\abstract{We study inclusive semileptonic $\bar{B}\to X_c \ell \nub$ decays ($\ell = e,\mu$) in the presence of generic physics Beyond the Standard Model (BSM) that is heavy compared to the $b$-quark mass.
Its effect is encoded in the Wilson coefficients of dimension-six operators of the Weak Effective Theory (WET). 
We compute the resulting BSM contributions to the experimentally measured kinematic moments and the total decay rate through order $\mathcal{O}(\LamQCD^3/m_b^3)$ in the Heavy Quark Expansion, as well as the $\mathcal{O}(\alpha_s)$ corrections for the right-handed vector current. 
We then perform fits to the full set of available inclusive measurements in several single-mediator scenarios and in the full WET basis.
We find qualitatively new bounds in several of these scenarios, complementary to and competitive with constraints from exclusive decays.}

\begin{document}
\allowdisplaybreaks

\maketitle

\section{Introduction}

Semileptonic $b \to c$ decays are the leading processes for the determination of the Cabibbo-Kobayashi-Maskawa (CKM) matrix element $V_{cb}$,
 a fundamental parameter of the Standard Model of particle physics (SM). $V_{cb}$ is  crucial for both a precise quantitative understanding of the SM and for indirect searches for physics Beyond the SM (BSM).
The latter statement holds true not only for charged-current processes, but also for flavour-changing neutral-current processes, which show a high sensitivity to $V_{cb}$.
As the QCD property of confinement prevents a direct measurement of $V_{cb}$, its determination always requires an interplay of data from experiment and theoretical calculations of the corresponding hadronic matrix elements.
From both theoretical and experimental points of view, processes with a specific hadronic final state (exclusive decays) are very different from processes that simply require a charm quark in the final state (inclusive decays).
These two classes therefore provide independent ways of determining $|V_{cb}|$ from $b \to c \ell \nub$ transitions.
Long-standing significant tensions between the exclusive and inclusive analyses motivated increasing efforts in both directions, see Refs.~%
\cite{Gambino:2020jvv,Bernlochner:2024sfg,Belle-II:2018jsg,HeavyFlavorAveragingGroupHFLAV:2024ctg,FlavourLatticeAveragingGroupFLAG:2024oxs} for recent reviews.

In this work we focus on inclusive decays to light leptons. The advantage of inclusive decays is that they allow for a local Operator Product Expansion (OPE), resulting in a 
double series in $\LamQCD/m_b$ and $\alpha_s$ known as the Heavy Quark Expansion (HQE).
Up to $\mathcal{O}(\LamQCD^3/m_b^3)$, the series only requires the knowledge of four nonperturbative parameters $\mu^2_\pi$, $\mu_G^2$, $\rho^3_D$, $\rho^3_{LS}$,  which can be extracted from sufficiently inclusive observables, such as the first few moments of the kinematic distributions. 
A state-of-the-art SM analysis of this kind recently achieved a relative accuracy for \Vcb of $1.1\%$~\cite{Finauri:2023kte,Bordone:2021oof,Fael:2020tow}, 
see also Ref.~\cite{Bernlochner:2022ucr}.
At this level of precision, it is natural to wonder about the implications for BSM physics.
Remaining agnostic regarding its specific form, effective field theories are the standard tool for tackling this type of problems.
We hence perform a model-independent study by taking dimension-six effective operators at the $B$ scale into account,
which arise from integrating out heavy degrees of freedom within and beyond the SM.

A difficulty arises from the fact that only few of the nonperturbative HQE parameters can be determined from inputs external to this analysis. 
This implies that the determination of the New Physics (NP) Wilson coefficients cannot be decoupled from the extraction of the HQE parameters, rendering their separate determination difficult. 
We therefore perform a global fit to the full set of inclusive $\bar{B} \to X_c \ell \nub$ observables, including external inputs where available, in order to simultaneously 
$(i)$ determine bounds on the magnitude of the NP Wilson coefficients in the Weak Effective Theory (WET) and 
$(ii)$ extract the pertinent HQE parameters. 
A promising way to obtain additional information on inclusive semileptonic decays and the corresponding HQE parameters is provided by  
lattice QCD calculations \cite{Gambino:2017vkx,Gambino:2020crt,DeSantis:2025yfm,Kellermann:2025pzt}. 
In time, they will allow to disentangle the determinations of BSM and HQE parameters.

BSM contributions to inclusive semileptonic $B$ decays have been previously studied in several publications, mostly devoted to the calculation of NP effects in the HQE~\cite{Czarnecki:1992zm,Grossman:1995yp,Dassinger:2007pj,Dassinger:2008as,Colangelo:2016ymy,Kamali:2018bdp,Colangelo:2020vhu,Fael:2022wfc}. 
A few phenomenological studies have also appeared~\cite{Feger:2010qc,Crivellin:2009sd,Crivellin:2014zpa,Celis:2016azn,Kamali:2018fhr,Jung:2018lfu,Iguro:2020cpg}, but they focus on specific scenarios (such as a right-handed vector current) or on the decays with a $\tau$ lepton.
With the exception of Ref.~\cite{Dassinger:2008as}, only the total decay rate has been used as a constraint on BSM physics, assuming the nonperturbative parameters obtained from the moments to be unaffected. 
A complete analysis implementing all relevant HQE corrections and performing a global fit to all available experimental data is still missing.
The purpose of this paper is to perform such an analysis. 

To that aim, we start by re-deriving the results of the existing literature. 
We largely agree with existing results, but identify a few points where our results are at variance with existing ones.
We calculate the impact of the new contributions to the full set of available inclusive observables, enabling us to present the first global fit to these observables, extracting simultaneously the HQE parameters and constraining up to seven additional NP degrees of freedom. 
We perform the fit first in the SM, to have a reference point for the impact of the BSM contributions. 
We then investigate specific BSM scenarios, motivated by the assumption of the presence of a weakly coupled single-mediator at the high scale~\cite{Freytsis:2015qca}. 
For this purpose we have studied the connection of the WET Wilson coefficients with those in the Standard Model effective field theory (SMEFT) at dimension-six.
Finally we perform the fit in the full WET with complex Wilson coefficients. 
We discuss in each case the implications for the inclusive extraction of $|V_{cb}|$.

The paper is structured as follows:
in Section~\ref{sec:theo} we outline the theoretical framework, comparing with the existing literature.
In Section~\ref{sec:numanal} we proceed with our numerical analysis, presenting the global fit results, and in Section~\ref{sec:conclusion} we present our conclusions.
The appendices contain additional tables and plots.

\section{Theoretical framework}
\label{sec:theo}
We start by presenting the calculation of the triple differential decay rate for $\bar{B} \to X_c \ell \nub$ in the presence of heavy NP. 
Assuming that physics beyond the SM exists at a scale $\LamNP \gg m_b$, the phenomenology of NP in $B$ meson decays can be performed in full generality in the WET framework.

At first order in WET, NP effects are encoded in Wilson coefficients multiplying dimension-six operators.
The weak effective Hamiltonian mediating $b \to c \ell \bar{\nu}_{\ell^\prime}$ processes is given by
\begin{equation}
\mathcal{H}_{\rm eff}= \frac{4 G_F}{\sqrt{2}} V_{c b} \sum_i \sum_{\ell^{\prime}} C_i^{\ell \ell^{\prime}} \mathcal{O}_i^{\ell \ell^{\prime}}+\text {h.c.}\,,
\end{equation}
where $G_F = 1.1663787\cdot 10^{-5}~\text{GeV}^{-2}$~\cite{ParticleDataGroup:2024cfk} is the Fermi constant
which we factored out together with $V_{cb}$ for convenience.
The dimension-six operators $\mathcal{O}_i^{\ell \ell^\prime}$ are constructed out of SM fermion fields and read
\begin{equation}
\begin{array}{ll}
\label{eq:NPopbasis}
\mathcal{O}_{V_{LX}}^{\ell \ell^{\prime}}= [\bar{c}\, \gamma^\mu P_L b] \; [\bar{\ell}\, \gamma_\mu P_X \nu_{\ell^{\prime}}]\,, &\qquad \mathcal{O}_{V_{RX}}^{\ell \ell^{\prime}} = [\bar{c}\, \gamma^\mu P_R b] \;[\bar{\ell}\, \gamma_\mu P_X \nu_{\ell^{\prime}}]\,, \\[2mm]
\mathcal{O}_{S_X}^{\ell \ell^{\prime}}= [\bar{c}\, b] \; [\bar{\ell}\, P_X \nu_{\ell^{\prime}}]\,, &\qquad \mathcal{O}_{P_X}^{\ell \ell^\prime} = [\bar{c}\, \gamma_5 b] \; [\bar{\ell}\, P_X \nu_{\ell^{\prime}}]\,, \\[2mm]
\mathcal{O}_{T_X}^{\ell \ell^{\prime}} = [\bar{c}\, \sigma^{\mu \nu} P_L b] \; [\bar{\ell}\, \sigma_{\mu \nu} P_X \nu_{\ell^{\prime}}]\,,
\end{array}
\end{equation}
with $P_{L,R} = (1\mp \gamma_5)/2$, $\sigma^{\mu\nu}=\frac{i}{2}[\gamma^\mu,\gamma^\nu]$, and $X = L,R$ selecting left-handed or right-handed neutrinos.
The Wilson coefficients $C_i^{\ell \ell^\prime}$ are dimensionless and encode the physics from scales larger than $m_b$.
The SM weak effective Hamiltonian is reproduced by $C_{V_{LL}}^{\ell \ell} = 1 + \mathcal{O}(\alpha_{\rm em})$, with all other Wilson coefficients set to zero. 
In the rest of this work we will assume lepton flavour conservation $\ell = \ell^\prime$ and left-handed neutrinos only ($X = L$).\footnote{If desired, it is straight-forward to relax these assumptions. For $\ell\neq \ell'$, the contributions are identical to those with $\ell=\ell'$, except that there is no interference with the latter. The contributions with $X=R$ again do not interfere with their $X=L$ counterparts and are also readily obtained from the ones provided here, using the transformations $C^{\ell \ell'}_{S_L}\to C^{\ell \ell'}_{S_R}$,  $C^{\ell \ell'}_{P_L} \to -C^{\ell \ell'}_{P_R}$, $C^{\ell \ell'}_{V_{LL}} \to C^{\ell \ell'}_{V_{RR}}$, $C^{\ell\ell'}_{V_{RL}}\to C^{\ell\ell'}_{V_{LR}}$, and $C^{\ell \ell'}_{T_L}\to C^{\ell \ell'}_{T_R}$.}
This leaves us with the SM operator $\mathcal{O}_{V_L}$ plus four additional NP operators, where we will drop the $\ell \ell$ superscript and the $L$ subscript related to neutrinos for simplicity.

In the literature the scalar and pseudoscalar operators are often replaced by the left and right-handed scalar operators $\mathcal{O}_{S_{L,R}} \equiv (\mathcal{O}_S \mp \mathcal{O}_P)/2$. We choose to work with the basis \eqref{eq:NPopbasis} instead because the interference between the scalar operator and pseudoscalar operator vanishes, due to parity conservation in QCD.

\subsection{Triple differential decay width}
The differential decay width is given by \cite{Blok:1993va,Manohar:1993qn}
\begin{align}
\label{eq:d3Gstart}
d\Gamma (\bar{B} \rightarrow X_c \ell \nub) &= \frac{1}{2m_B} \frac{d^3 p_\ell}{\left(2\pi \right)^3 2E_\ell} \frac{d^3 p_\nub}{\left(2\pi \right)^3 2E_{\nub}}  \sum_{X_c} (2\pi)^4 \delta^4\left(p_B - p_{X_c} - p_\ell -p_{\nub} \right)\nonumber\\
&\quad\times\sum_{\text{spins}}\left| \langle X_c \,\ell \, \nub \left | \mathcal{H}_\mathrm{eff} \right| \bar{B} \rangle \right|^2 \nonumber\\
&= \;64\pi G_F^2 |V_{cb}|^2 \frac{d^3 p_\ell}{\left(2\pi \right)^3 2E_\ell} \frac{d^3 p_\nub}{\left(2\pi \right)^3 2E_{\nub}} \sum_{\Gamma,\Gamma'} L_{\Gamma,\Gamma'} W_{\Gamma,\Gamma'}\,,
\end{align}
where the sum over $X_c$ states includes the phase space integral for the final state hadronic system.
In the last line the sum on $\Gamma,\Gamma^\prime$ runs over $\{\mathds{1}, \gamma^\mu, \sigma^{\mu\nu} \}$, with the leptonic and hadronic tensors defined as
\begin{align}
\label{eq:lepthadrtensDef}
L_{\Gamma,\Gamma'} &= \frac{1}{4} \sum_{\text{spins}} \left\langle \nub \ell \left| \bar{\ell}\, \Gamma P_L \,\nu_\ell \right| 0 \right\rangle \left\langle \nub \ell \left| \bar{\ell} \,\Gamma' P_L\, \nu_\ell \right| 0 \right\rangle^\dagger = \frac{1}{4}\text{tr}[(\slashed{p}_\ell + m_\ell)\Gamma P_L\slashed{p}_{\nub}\Gamma^\prime]\,,\nonumber \\
W_{\Gamma,\Gamma'} &= \frac{\left(2\pi \right)^3}{2m_B} \sum_{X_\mathrm{c}} \delta^4 (p_B - p_{X_c} - q) \left\langle X_c \left | \bar{c} \,\Gamma d_\Gamma \, b \right| \bar{B} \right\rangle \left\langle X_c \left|\bar{c} \,\Gamma' d_{\Gamma^\prime}\, b \right| \bar{B} \right\rangle^\dagger\,,
\end{align}
where we defined $q = p_\ell + p_\nub$, and the following matrices in spinor space
\begin{equation}
    d_{\gamma_\mu} = C_{V_L} P_L + C_{V_R} P_R\,, \qquad d_{\mathds{1}} = C_S + C_P \gamma_5\,, \qquad d_{\sigma_{\mu \nu}} = C_T P_L\,.
\end{equation}
From the definitions~\eqref{eq:lepthadrtensDef} we obtain the useful relations $L_{\Gamma',\Gamma} = L^*_{\Gamma,\Gamma'}$ and $W_{\Gamma',\Gamma} = W^*_{\Gamma,\Gamma'}$.
The leptonic tensors can be computed as functions of the four-vectors $p_\ell$ and $q$:
\begin{align}
\label{eq:LGG}
L_{\mathds{1},\mathds{1}} &= \frac{1}{4}(q^2 - m_\ell^2)\,, \nonumber \\
L_{\gamma^\mu, \gamma^\nu} &=  -\frac{g^{\mu\nu}}{4} (q^2 - m_\ell^2) - p_\ell^\mu p_\ell^\nu +\frac{1}{2}\left(p_\ell^\mu q^\nu + p_\ell^\nu q^\mu \right) - \frac{i}{2} \epsilon^{\mu\nu\alpha\beta} {p_\ell}_\alpha q_\beta\,, \nonumber\\
L_{\sigma^{\kappa\lambda},\sigma^{\rho \tau}} &= \frac{1}{4} \biggl[ \left(g^{\kappa\tau} g^{\lambda \rho} - g^{\kappa\rho}g^{\lambda\tau} \right) \left(m_\ell^2 - q^2 \right) \nonumber \\
& + 4 p_\ell^\kappa \left(p_\ell^\rho g^{\lambda \tau} -p_\ell^\tau g^{\lambda\rho} \right) + 4 p_\ell^\lambda \left(p_\ell^\tau g^{\kappa\rho} - p_\ell^\rho g^{\kappa\tau} \right) \nonumber \\
&+ 2g^{\kappa \tau} \left(p_\ell^\lambda q^\rho + p_\ell^\rho q^\lambda \right) - 2g^{\lambda\tau}\left(p_\ell^\kappa q^\rho + p_\ell^\rho q^\kappa \right) \nonumber \\
& - 2g^{\kappa\rho} \left(p_\ell^\lambda q^\tau + p_\ell^\tau q^\lambda \right) + 2g^{\lambda \rho} \left(p_\ell^\kappa q^\tau + p_\ell^\tau q^\kappa \right)\biggr]  \nonumber \\
& + \frac{i}{2} \biggl[ \left(p_\ell^\lambda \epsilon^{\kappa\rho\tau\alpha} -p_\ell^\kappa \epsilon^{\lambda\rho\tau\alpha} + p_\ell^\rho \epsilon^{\kappa\lambda\tau\alpha} - p_\ell^\tau \epsilon^{\kappa\lambda\rho\alpha} \right) {p_\ell}_\alpha \nonumber \\
& \phantom{+ 2 i \Big[ (} + \left(q^\tau \epsilon^{\kappa\lambda\rho\alpha} -q^\rho \epsilon^{\kappa\lambda\tau\alpha} \right) {p_\ell}_\alpha + \left(p_\ell^\kappa \epsilon^{\lambda\rho\tau\alpha} - p_\ell^\lambda \epsilon^{\kappa\rho\tau\alpha} \right)q_\alpha \biggr]\,,  \nonumber \\
L_{\mathds{1}, \gamma^\mu} &= \frac{m_\ell}{2} (q^\mu - p_\ell^\mu) \,,\nonumber \\
L_{\mathds{1}, \sigma^{\kappa\lambda}} &= \frac{1}{2} \left[i \left(p_\ell^\lambda q^\kappa - p_\ell^\kappa q^\lambda \right) + \epsilon^{\kappa\lambda\alpha\beta} {p_\ell}_\alpha q_\beta \right]\,,\nonumber \\
L_{\gamma^\mu, \sigma^{\kappa\lambda}} &= \frac{m_\ell}{2} \Bigl[ i  \left(g^{\mu\kappa} p_\ell^\lambda - g^{\mu\lambda} p_\ell^\kappa + g^{\mu\lambda} q^\kappa - g^{\mu\kappa} q^\lambda \right) + \epsilon^{\kappa\lambda\mu\alpha}(q_\alpha - {p_\ell}_\alpha) \Bigr]\,, 
\end{align}
where we use the convention $\epsilon^{0123} = +1$ for the Levi-Civita fully antisymmetric tensor.
The hadronic tensor can depend only on two vectors, the normalized dilepton momentum $\hat{q}^\mu = q^\mu/m_b$ and the heavy meson velocity $v^\mu = p_B^\mu/m_B$ (with $v^2=1$).
Before computing the $W_{\Gamma,\Gamma'}$ we hence parametrize them in terms of dimensionless scalar functions $w_i(\hat{q}^2,\hat{q}_0=\hat{q}\cdot v)$ by decomposing the tensor structures in a minimal basis\footnote{To find the minimal decomposition we employ the Schouten's identity (for any four-vector $V^\mu$):
\[
V^\tau \epsilon^{\mu\nu\rho\sigma} + V^\mu \epsilon^{\nu\rho\sigma\tau} + V^\nu \epsilon^{\rho\sigma\tau\mu} + V^\rho \epsilon^{\sigma\tau\mu\nu} + V^\sigma \epsilon^{\tau\mu\nu\rho} = 0\,.  
\]}.
In addition we note that the Dirac matrix inside the tensor current can be written as $\sigma^{\mu\nu}P_L = {\Sigma^{\mu\nu}}_{\alpha\beta}\sigma^{\alpha\beta}$, with the projector
\begin{equation}
    \Sigma^{\mu\nu\alpha\beta} = \frac{1}{4}\biggl(g^{\mu\alpha}g^{\nu\beta} - g^{\mu\beta}g^{\nu\alpha} - i \epsilon^{\mu\nu\alpha\beta} \biggr)\,,
\end{equation}
which constrains the Lorentz structure decomposition.
To further simplify the form of the hadronic tensor we exploit the symmetry of QCD under time reversal $T$, obtaining the condition
\begin{equation}
\label{eq:Tconst}
    W_{\Gamma,\Gamma'}(\hat{q},v) = s_\Gamma s_{\Gamma'} W_{\Gamma,\Gamma'}^*(\tilde{\hat{q}},\tilde{v})|_{C_i \to C_i^*}\,.
\end{equation}
Here $s_\Gamma$ is the sign induced by the transformation specific to each current
\begin{equation}
    s_{\mathds{1}} = 1\,, \qquad s_{\gamma^\mu} = \tilde{I}^\mu\,, \qquad s_{\sigma^{\mu\nu}} = -\tilde{I}^\mu\tilde{I}^\nu\,,
\end{equation}
with $\tilde{I}^\mu = (1,-1,-1,-1)$ and the parity transformed vectors are $\tilde{V}^\mu = (V^0, -\vec{V})$ for $V^\mu = \{\hat{q}^\mu,v^\mu\}$.
By using the fact that (no summation over indices)
\begin{equation}
    \tilde{I}^\mu \tilde{V}^\mu = V^\mu\,,\qquad \tilde{I}^\mu \tilde{I}^\nu g^{\mu\nu} = g^{\mu\nu}\,, \qquad \tilde{I}^\mu \tilde{I}^\nu \tilde{I}^\rho \tilde{I}^\sigma \epsilon^{\mu\nu\rho\sigma} = - \epsilon^{\mu\nu\rho\sigma}\,,
\end{equation}
and combining~\eqref{eq:Tconst} with $W^*_{\Gamma,\Gamma'} = W_{\Gamma',\Gamma}$ we can write
\begin{align}
\label{eq:WGG}
m_b W_{\mathds{1},\mathds{1}} &= w_0\,,\nonumber\\
m_b W_{\gamma^\mu, \gamma^\nu} &= -g^{\mu\nu} w_1+v^{\mu}v^{\nu}w_2+i\epsilon^{\mu\nu\alpha\beta}v_\alpha \hat q_\beta
w_3+\hat q^\mu \hat q^\nu w_4+(\hat q^\mu v^{\nu}+\hat q^\nu v^{\mu})w_5\,,\nonumber\\
m_b W_{\sigma^{\mu\nu}, \sigma^{\alpha\beta}} &= {\Sigma^{\mu\nu}}_{\kappa\lambda} {\Sigma^{*\alpha\beta}}_{\rho\tau}\biggl\{\left[v^\kappa\left(v^\tau g^{\lambda\rho}-v^\rho g^{\lambda\tau}\right)-v^\lambda\left(v^\tau
g^{\kappa\rho}-v^\rho g^{\kappa\tau}\right)\right]w_6 \nonumber \\ 
&\phantom{=}+\left[\hat q^\kappa\left(\hat q^\tau g^{\lambda\rho}-\hat
q^\rho g^{\lambda\tau}\right)-\hat q^\lambda\left(\hat q^\tau g^{\kappa\rho}-\hat q^\rho g^{\kappa\tau}\right)\right]w_7\nonumber\\
&\phantom{=}+\Big[g^{\kappa\tau}\left(v^\lambda \hat q^\rho+v^\rho\hat
q^\lambda\right)-g^{\lambda\tau}\left(v^\kappa \hat q^\rho+v^\rho\hat q^\kappa\right) \nonumber \\
& \phantom{=+} -g^{\kappa\rho}\left(v^\lambda \hat q^\tau+v^\tau\hat q^\lambda\right)+g^{\lambda\rho}\left(v^\kappa \hat q^\tau+v^\tau\hat q^\kappa\right)\Big]w_{8}\biggr\}\,,\nonumber\\
m_b W_{\mathds{1}, \gamma^\mu} &= \hat q^\mu w_9 + v^{\mu} w_{10}\,,\nonumber\\
m_bW_{\mathds{1}, \sigma^{\rho \tau}} &= \frac{i}{2}(v^{\rho}\hat{q}^\tau-v^{\tau}\hat{q}^\rho + i\epsilon^{\rho\tau\alpha\beta}v_\alpha \hat{q}_\beta)w_{11}\,,\nonumber\\
m_b W_{\gamma^\mu, \sigma^{\rho \tau}} &=
\frac{i}{2}(g^{\rho\mu}v^{\tau}-g^{\tau\mu}v^{\rho} +i \epsilon^{\mu\rho\tau\alpha}v_\alpha)w_{12}+\frac{i}{2}(g^{\rho\mu}\hat{q}^{\tau}-g^{\tau\mu}\hat{q}^{\rho} +i \epsilon^{\mu\rho\tau\alpha}\hat{q}_\alpha)w_{13}\nonumber\\
&+\frac{i}{2}(v^{\rho}\hat{q}^\tau-v^{\tau}\hat{q}^\rho + i \epsilon^{\rho\tau\alpha\beta}v_\alpha \hat{q}_\beta)(v^{\mu}w_{14}+\hat{q}^\mu w_{15})\,,
\end{align}
In this convention the factors of $i$ (the imaginary unit) are introduced such that the $w_i$'s are real for real Wilson coefficients.
The constraint~\eqref{eq:Tconst} prevents in particular the presence of a term $(v^\mu \hat{q}^\nu - v^\nu \hat{q}^\mu)$ in $W_{\gamma^\mu,\gamma^\nu}$ and of a term $\epsilon^{\mu\nu\alpha\beta}$ in $W_{\sigma^{\mu\nu},\sigma^{\alpha\beta}}$.

We can now evaluate $\sum_{\Gamma,\Gamma'} L_{\Gamma,\Gamma'}W_{\Gamma,\Gamma'}$, by using Eqs.~\eqref{eq:LGG} and~\eqref{eq:WGG}, and insert it into Eq.~\eqref{eq:d3Gstart}  
\begin{align}
\label{eq:d3Gml}
\frac{d^3 \Gamma}{d \hat{q}^2 d \hat{E}_{\ell} d \hat{E}_\nub} &= \frac{G_F^2|V_{cb}|^2 m_b^5}{\pi^3}\theta(\hat{E}_\nub)\theta(\hat{E}_\ell-\hat{m}_\ell) \theta\Bigl(\hat{q}^2-2\hat{E}_\nub(\hat{E}_\ell-\sqrt{\hat{E}_\ell^2-\hat{m}_\ell^2})-\hat{m}^2_\ell\Bigr)\nonumber\\
&\times\theta\Bigl(2\hat{E}_\nub(\hat{E}_\ell + \sqrt{\hat{E}_\ell^2-\hat{m}_\ell^2}) + \hat{m}^2_\ell - \hat{q}^2\Bigr)\biggl\{(w_0+2w_1+\hat{m}_\ell^2 w_4+2\hat{m}_\ell w_{9_R})\frac{\hat{q}^2-\hat{m}_\ell^2}{4}\nonumber\\
&+\frac{w_2}{4} (4 \hat{E}_\ell \hat{E}_\nub+\hat{m}_\ell^2-\hat{q}^2)+\left(\frac{w_3}{2}- w_{11_R}-\hat{m}_\ell w_{15_R}\right) \left(\hat{E}_\ell (\hat{m}_\ell^2-\hat{q}^2)+\hat{E}_\nub (\hat{m}_\ell^2+\hat{q}^2)\right)\nonumber\\
&+\hat{E}_\nub \hat{m}_\ell (\hat{m}_\ell w_{5} + w_{10_R} + 6w_{12_R}) +w_{6} (8 \hat{E}_\ell \hat{E}_\nub+\hat{m}_\ell^2-\hat{q}^2) +w_{7} (\hat{q}^4 +\hat{m}_\ell^2 \hat{q}^2 -2 \hat{m}_\ell^4)\nonumber\\
&+2 w_{8} \left( (\hat{E}_\ell+E_\nub) \hat{q}^2-\hat{E}_\ell \hat{m}_\ell^2  + 3\hat{E}_\nub \hat{m}_\ell^2 \right) + 3\hat{m}_\ell w_{13_R}(\hat{q}^2-\hat{m}_\ell^2) \nonumber\\
&+\hat{m}_\ell w_{{14}_R} \left(\hat{q}^2-2 \hat{E}_\nub (\hat{E}_\ell+\hat{E}_\nub)-\hat{m}_\ell^2\right)
\biggr\}\,,
\end{align}
where we performed the trivial angular integrations and defined normalized quantities $\hat{m}_\ell = m_\ell/m_b$, $\hat{E}_\ell = E_\ell/m_b$, $\hat{E}_\nub = E_\nub/m_b$.
In the interference terms we employed the short-hand  notation $w_{i_R}= \textrm{Re}(w_i)$.

In the following we will neglect the lepton mass, as we are only interested in electrons and muons.
In this case, changing variables with $\hat{E}_{\nub} = \hat{q}_0 - \hat{E}_\ell$,~\eqref{eq:d3Gml} simplifies to
\begin{align}
\label{eq:d3Gnoml}
\frac{d^3 \Gamma}{d \hat{q}^2 d \hat{E}_{\ell} d \hat{q}_0} &= \frac{G_F^2|V_{cb}|^2 m_b^5}{\pi^3}\theta(\hat{q}^2)\theta(\hat{E}_\ell) \theta\Bigl(4(\hat{q}_0 - \hat{E}_\ell) \hat{E}_\ell - \hat{q}^2\Bigr)\biggl\{(w_0+2w_1)\frac{\hat{q}^2}{4}\nonumber\\
&+\frac{w_2}{4} \left(4 \hat{E}_\ell (\hat{q}_0-\hat{E}_\ell)-\hat{q}^2\right)+\left(\frac{w_3}{2}-w_{11_R}\right)(\hat{q}_0-2\hat{E}_\ell) \hat{q}^2\nonumber\\
&+w_{6} \left(8 \hat{E}_\ell (\hat{q}_0 - \hat{E}_\ell)-\hat{q}^2\right) +w_{7} \hat{q}^4 +2 w_{8} \hat{q}_0 \hat{q}^2 \biggr\}\,.
\end{align}
We immediately notice that only a subset of the structure functions $w_i$ from~\eqref{eq:WGG} contributes to the massless lepton case.
In particular we see that interference terms of the form $\text{Re}[C_P^* C_{V_X}]$, $\text{Re}[C_S^* C_{V_X}]$ ($w_{9,10}$), and $\text{Re}[C_T^* C_{V_X}]$ ($w_{12-15}$) are not present.
This means that the only possible interference term between SM and NP is given by $\text{Re}[C_{V_L} C_{V_R}^*]$ in $w_{1-5}$. Moreover, as anticipated above, the interference between the scalar and pseudoscalar operators in $w_0$ is also absent because of parity conservation.

It is useful to introduce the charm quark off-shellness
\begin{equation}
\hat{u} = \frac{(m_b v-q)^2-m_c^2}{m_b^2}\,,    
\end{equation}
and the normalized squared charm mass 
\begin{equation}
    \rho = \frac{m_c^2}{m_b^2}\,,
\end{equation}
in terms of which
\begin{equation}
\hat{q}_0 = \frac12(1-\rho +\hat{q}^2 - \hat{u})\,.    
\end{equation}
In this way the structure functions only depend on $\hat{q}^2$ and $\hat{u}$, while the $\hat{E}_\ell$ dependence of the triple differential decay rate is fully determined.
It is then possible to obtain the double differential decay rate by integrating over the whole $E_\ell$ allowed phase space without determining the structure functions $w_i(\hat{q}^2,\hat{u})$:
\begin{align}
\label{eq:d2Gnoml}
\frac{d^2 \Gamma}{d \hat{q}^2 d\hat{u}} &= \frac{G_F^2|V_{cb}|^2 m_b^5}{24\pi^3}\theta(\hat{q}^2) \theta\left((1-\sqrt{\hat{q}^2})^2-\rho-\hat{u}\right)\sqrt{\hat{q}_0^2-\hat{q}^2}\nonumber\\
&\times\biggl\{2 \hat{q}_0^2 (w_{2}+8 w_{6})+\hat{q}^2 (3w_{0}+6 w_{1}-2 w_{2} +24 \hat{q}_0 w_{8}-4 w_{6})+12 \hat{q}^4 w_{7} \biggr\}\,,
\end{align}
where $\hat{q}_0$ has to be considered as a function of $\hat{q}^2$ and $\hat{u}$.
Upon integration over the whole lepton energy range, the contributions from $w_3$ and $w_{11}$ vanish. 
As a consequence, the number of independent SM structures is reduced to two, and interference terms of the form $\text{Re}[C_T^*C_S]$ and $\text{Re}[C_T^*C_P]$ do not contribute to the total width or the $q^2$ moments, as we will see in Section~\ref{sec:observables}.

The forward-backward asymmetry represents an additional and potentially orthogonal observable~\cite{Turczyk:2016kjf,Fael:2022wfc, Herren:2022spb}. If one uses a lower cut on $q^2$, as suggested in Ref.~\cite{Herren:2022spb}, this asymmetry can be written as
\begin{equation}
\label{eq:AFB}
    A_{FB}(q^2_{\rm cut}) = \frac{G_F^2|V_{cb}|^2 m_b^5}{8\pi^3 \Gamma|_{q^2>q^2_{\rm cut}}}\int_{\hat{q}^2_{\rm cut}}^\infty d\hat{q}^2 \,\hat{q}^2 \int d\hat{u} \,\theta\Bigl((1-\sqrt{\hat{q}^2})^2-\rho-\hat{u}\Bigr)(\hat{q}_0^2-\hat{q}^2)(2 w_{11_R}-w_3)\,,
\end{equation}
and is sensitive precisely to the structure functions that do not contribute to the $q^2$ moments. 
This makes its future measurement at Belle II very interesting because it adds sensitivity to both $\mu^2_G$ and NP.

\subsection{Local OPE of the hadronic tensor with New Physics}
The hadronic tensor can be calculated thanks to the optical theorem and an OPE of the forward scattering amplitude for the $\bar{B}$ meson,
\begin{equation}
\label{eq:TDef}
T_{\Gamma,\Gamma'} = -\frac{i}{2m_B} \int d^4x e^{-iq\cdot x}\langle \bar{B} | \text{T}\{J_{\Gamma'}^\dagger(x) J_{\Gamma}(0)\} | \bar{B} \rangle \,.
\end{equation}
The forward scattering amplitude has the same structure function decomposition as the hadronic tensor~\eqref{eq:WGG}, and the two sets  
of coefficients are related by\footnote{When the Wilson coefficients are real, this relation reduces to the usual relation $w_i = -\frac{1}{\pi} \textrm{Im} [t_i]$.}
\begin{equation}
w_i = \frac{i}{2\pi} \textrm{disc} [t_i]\,,
\end{equation}
where $t_i$ are the structure functions of the forward scattering amplitude and the discontinuity refers to the variable $\hat{q}_0$. 
We first compute $T_{\Gamma,\Gamma'}$ at tree level up to $\mathcal{O}(\LamQCD^3/m_b^3)$ using the background field method for the charm quark propagator~\cite{Blok:1993va,Novikov:1984ecy}. Our results, after considering the change of tensor basis, are in agreement with the results from Ref.~\cite{Colangelo:2020vhu}.
Moreover we calculate $\mathcal{O}(\alpha_s)$ perturbative effects at leading power in the NP terms containing $C_{V_R}$~\cite{Dassinger:2007pj,Alberti:2015qmj}, as it is the only contribution having a non-vanishing interference with the SM. 
Indeed, that interference is the only NP contribution to the decay distributions that is linear in the NP Wilson coefficients.
The rest of the NP operators contribute only quadratically, and might be more suppressed with respect to the SM.
We provide the expressions for the $w_i$'s in a Mathematica package file $\texttt{wiNP.m}$ attached to this preprint.

We have successfully checked our results for NP contributions against available results in the literature. These checks are listed in Table~\ref{tab:NPchecks}. However, we only partially agree with the numerical results of Ref.~\cite{Fael:2022wfc}, where the complete $\mathcal{O}(\LamQCD^3/m_b^3)$ and $\mathcal{O}(\alpha_s)$ corrections have been considered. We believe that part of the discrepancy originates from a problem in the re-expansion of normalized observables in powers of small Wilson coefficients (Eq.\ (26) of Ref.~\cite{Fael:2022wfc}).\footnote{We thank the authors of Ref.~\cite{Fael:2022wfc} for confirming the presence of a few typos and missing factors of 2 in their results.}
The same issues arise when we compare our results with the python package \texttt{kolya}~\cite{Fael:2024fkt}.

\begin{table}[]
    \centering
    \begin{tabular}{c|c|c|c}
         Observable & NP & Order & Ref. \\ \hline
         $\Gamma_{\rm tot}$ & all & LO &  \cite{Jung:2018lfu} \\ 
         $d\Gamma/d\hat q^2$ & $C_{S,P}$ & LO & \cite{Celis:2016azn} \\         
         $d\Gamma/d\hat q^2$ & all & $\mathcal{O}(\LamQCD^2/m_b^2)$ & \cite{Kamali:2018bdp} \\
         $d\Gamma/d\hat{E}_\ell$ & $C_{S_R}$ & $\mathcal{O}(\LamQCD^2/m_b^2)$& \cite{Grossman:1995yp} \\
          $d\Gamma/d\hat{q}^2$ & $C_T$ &  $\mathcal{O}(\LamQCD^2/m_b^2)$ & \cite{Colangelo:2016ymy} \\
          $\langle E_\ell^n\rangle$ & $C_{V_R} $ &  $\mathcal{O}(\LamQCD^2/m_b^2,\alpha_s)$ &  \cite{Dassinger:2008as} \\ 
          $d^3 \Gamma / d\hat q^2 d \hat E_\ell d \hat q_0$ & all & $\mathcal{O}(\LamQCD^3/m_b^3)$ & \cite{Colangelo:2020vhu}
    \end{tabular}
    \caption{Cross-check with existing calculations for NP contributions. Some of the references (last column) only consider specific operators, specified in the ``NP'' column.}
    \label{tab:NPchecks}
\end{table}
\setlength{\tabcolsep}{1pt} 
\subsection{Parametrization of New Physics effects}
The WET Wilson coefficients are complex quantities, which would imply in principle 10 free parameters.
However, there is no way of distinguishing the NP effects in the SM-like contribution $C_{V_L}$ in a single process, as this coefficient can be absorbed into $V_{cb}$.
We hence define\footnote{Here $C_{V_L}^{\rm NP}$ stands for the NP part of $C_{V_L}$. The universal electroweak corrections will be factored out explicitly as usually done in the literature.} $\tilde{V}_{cb} = \bigl(1+C_{V_L}^{\rm NP}\bigr) V_{cb}$ and redefine the rest of the Wilson coefficients as
\begin{equation}
    \tilde{C}_i = \frac{C_i}{C_{V_L}}\,.
\end{equation}
The fully differential decay rate is proportional to $|\tilde{V}_{cb}|^2$, hence we will only be able to determine  the modulus of $\tilde{V}_{cb}$.
As noticed below Eq.~\eqref{eq:d3Gnoml}, with the basis choice~\eqref{eq:NPopbasis}, not all interference terms appear in the differential decay rate when neglecting the lepton mass. 
This allows us to further reduce the number of free parameters.
In particular there are only three non-vanishing interference terms
\begin{align}
   \frac{1}{|C_{V_L}|^2} \text{Re}[C_{V_L}^* C_{V_R}] &= \text{Re}[\tilde{C}_{V_R}] = a_R\, \cos(\delta_R)  \,,\nonumber\\
   \frac{1}{|C_{V_L}|^2} \text{Re}[C_S C_T^*] &= \text{Re}[\tilde{C}_S \tilde{C}_T^*] = a_S \,a_T \,\cos(\delta_{ST})\,, \nonumber\\
   \frac{1}{|C_{V_L}|^2} \text{Re}[C_P C_T^*] &= \text{Re}[\tilde{C}_P \tilde{C}_T^*] = a_P \,a_T \,\cos(\delta_{PT})\,,
\end{align}
giving us access to the cosine of three relative phases $\delta_{R} \equiv \textrm{Arg}(\tilde{C}_{V_R})$ and $\delta_{XT} \equiv \textrm{Arg}(\tilde{C}_X \tilde{C}_T^*)$, and the absolute values $a_{S,P,T,R} \equiv |\tilde{C}_{S,P,T,V_R}|$.
The NP effects for inclusive semileptonic $B$ decays are hence parametrized by the following set of seven independent parameters:
\begin{equation}
    \{a_R\,, a_S\,, a_P\,, a_T\,, \cos(\delta_R)\,, \cos(\delta_{ST})\,, \cos(\delta_{PT})\}\,,
\end{equation}
with $a_i \geq 0$ and $-1\leq \cos(\delta_i)\leq 1$.

\subsection{Single mediator scenarios}
We conclude this section by introducing a few plausible minimal BSM scenarios. 
In these scenarios we assume the existence of a single new mediator with a mass of order $\Lambda_{\textrm{NP}}\gg m_W$ weakly coupled to the SM. 
At energy scales below $\LamNP$ the phenomenology of these scenarios can be studied by computing their contributions to the SMEFT Lagrangian at dimension six.
Conveniently, each scenario contributes to a small and unique set of SMEFT operators which acts as a phenomenological signature of the model. 
Since our observables are calculated at the $B$ mass scale, we will need to relate the Wilson coefficients generated at $\LamNP$ to the WET Wilson coefficients evolved down to the scale $\mu \simeq m_B$. 
The part of the SMEFT Lagrangian  we are interested in is
\begin{align}
\label{eq:LSMEFT}
    \mathcal{L}_{\rm SMEFT} \supset \frac{1}{\LamNP^2}\sum_{i=1}^3\biggl[& [\hat{C}_{V_L}]_i [\mathcal{Q}_{\ell q}^{(3)}]_i +[\hat{C}_{V_R}]_i[\mathcal{Q}_{\varphi ud}]_i+ [\hat{C}_{S_L}]_i [\mathcal{Q}_{\ell equ}^{(1)}]^\dagger_i \nonumber \\
    & +[\hat{C}_{S_R}]_i [\mathcal{Q}_{\ell edq}]^\dagger_i +[\hat{C}_T]_i [\mathcal{Q}_{\ell equ}^{(3)}]^\dagger_i  \biggr]\,,
\end{align}
where $i$ is the flavour index related to the field which will give the down-type quarks in the broken phase of the SM. 
We implicitly sum over light lepton flavours (assuming lepton flavour conservation), while we fix the flavour of the up-type quarks in order to select operators coupling to the charm quark.
The dimension-six operators $\mathcal{Q}_k$ in Eq.~\eqref{eq:LSMEFT} follow the notation of the Warsaw basis~\cite{Grzadkowski:2010es} and are written in terms of the SMEFT fermion fields $L, Q, E, D, U$, corresponding to the left-handed lepton doublet, the left-handed quark doublet and the right-handed lepton, down-quark and up-quark singlets, respectively.
For the Wilson coefficients and operators we chose the flavour basis where the Yukawa coupling to the $U$ fields is diagonal (up-aligned basis).
The explicit form of the operators is
\begin{align} 
\label{eq:SMEFToperators}
    [\mathcal{Q}_{\ell q}^{(3)}]_i &= [\bar{L} \gamma_\mu \tau^I L]\, [\bar{Q}_2 \gamma^\mu \tau^I Q_i]\,,\nonumber\\
    [\mathcal{Q}_{\varphi u d}]_i &= \tilde{\varphi}^\dagger i D_\mu \varphi [\bar{U}_2 \gamma^\mu D_i]\,,\nonumber\\
    [\mathcal{Q}_{\ell equ}^{(1)}]_i &= \epsilon_{jk}[\bar{L}^j E]\,[\bar{Q}^k_i U_2]\,,\nonumber\\
    [\mathcal{Q}_{\ell edq}]_i &= [\bar{L}^j E]\, [\bar{D}_i Q^j_2]\,,\nonumber\\
    [\mathcal{Q}_{\ell equ}^{(3)}]_i &= \epsilon_{jk}[\bar{L}^j \sigma_{\mu\nu} E]\,[\bar{Q}^k_i \sigma^{\mu\nu} U_2]\,,
\end{align}
where the superscripts on $L$ and $Q$ denote the SU(2)$_L$ indices, $\epsilon_{12} = 1$ and $\tau^I$ are the Pauli matrices (implicit sum over $I=1,2,3$). 
The subscripts on quark fields are generation indices, which are kept implicit on lepton fields.
The Higgs doublet $\varphi$ also appears as $\tilde{\varphi}_j = \epsilon_{jk} \varphi_k^*$.
It is worth mentioning that the basis~\eqref{eq:SMEFToperators} includes all the SMEFT operators generating, up to the one-loop level, non-vanishing matching onto~$\mathcal{O}_{V_R}$, $\mathcal{O}_S$, $\mathcal{O}_P$, $\mathcal{O}_T$ in WET, while there are other operators contributing to $\mathcal{O}_{V_L}$ which at the $B$ scale we cannot distinguish from the contribution of $\mathcal Q_{\ell q}^{(3)}$.

We employ the \texttt{Mathematica} package \texttt{DsixTools}~\cite{Celis:2017hod,Fuentes-Martin:2020zaz} to perform the running within SMEFT from $\LamNP$ to the electroweak scale, the one-loop matching between SMEFT and WET at the electroweak scale, the rotation to the mass basis for down-type quarks, and finally the running within WET to the low scale $\mu = 5~\text{GeV}$. At the level of one-loop matching, the SMEFT operators listed in Eq.~\eqref{eq:SMEFToperators} contribute to the WET operators relevant to the phenomenology of $b\to c\ell \nub$ decays listed in Eq.~\eqref{eq:NPopbasis}, as well as to other operators relevant for different low-energy observables. In this work, we choose to focus exclusively on the phenomenology of inclusive $b\to c\ell \nub$ decays. A global fit of SMEFT Wilson coefficients is beyond the scope of this article. The WET Wilson coefficients, at $\mu = 5~\text{GeV}$, are given by the following linear combinations of SMEFT Wilson coefficients at $\LamNP = 1$~TeV  (see also~\cite{Murgui:2019czp})
\begin{align}
\label{eq:WCrun}
    C_{V_L} &= 1 -0.0655 [\hat{C}_{V_L}]_2 - 1.52 [\hat{C}_{V_L}]_3\,,\nonumber\\
    C_{V_R} &= 0.689 [\hat{C}_{V_R}]_3\,, \nonumber\\
    C_{S_L} &= (-0.012+0.0038 i) [\hat{C}_{S_L}]_1 -0.0485 [\hat{C}_{S_L}]_2 - 1.13[\hat{C}_{S_L}]_3 + 0.008[\hat{C}_{T}]_2 + 0.158[\hat{C}_{T}]_3\,,\nonumber\\
    C_{S_R} &= -1.13[\hat{C}_{S_R}]_3\,,\nonumber\\
    C_T \; \, &= 0.0032[\hat{C}_{S_L}]_3 +0.0021i [\hat{C}_T]_1- 0.0275 [\hat{C}_{T}]_2 -0.638 [\hat{C}_{T}]_3\,,
\end{align}
where terms with coefficients smaller than $10^{-3}$ have been neglected. 
Due to the structure of the SMEFT operators we have used here the combinations
\begin{align}
    C_{S_L} = C_S-C_P\,, \qquad C_{S_R} = C_S + C_P\,.
\end{align}

We now select the tree-level single-mediator models in which the matching between the UV completion and SMEFT generates the $\hat{C}_i$ at the $\LamNP$ scale.
We consider a vector-like quark doublet, a charged scalar $H^\pm$, and coloured scalar ($S_1$, $R_2$) or vector ($U_1,V_2$) leptoquarks.
They can be grouped in five categories~\cite{Freytsis:2015qca}:
\begin{align}
\label{eq:treeUVmodels}
    (\text{Vector-like doublet}) &\to \{\hat{C}_{V_R}\}\,, \qquad \qquad \qquad\;\;\, (\text{I})\nonumber\\
    (H^{\pm})&\to\{\hat{C}_{S_L},\hat{C}_{S_R}\}\,, \qquad\qquad\;\, (\text{II})\nonumber\\
    (S_1) &\to \{\hat{C}_T, \hat{C}_{S_L} = -4\hat{C}_T\}\,, \;\,\,\, (\text{III})\nonumber\\
    (R_2) &\to \{\hat{C}_T, \hat{C}_{S_L} = 4\hat{C}_T\}\,,\;\;\quad (\text{IV})\nonumber\\
    (U_1,V_2) &\to \{\hat{C}_{S_R}\}\,,\qquad\qquad\qquad\;\, (\text{V})
\end{align}
where we do not consider whether $\hat{C}_{V_L}$ is generated as well since, according to Eq.~\eqref{eq:WCrun}, it would only induce a non-vanishing $C_{V_L}$ in WET, to which we are not sensitive.
By plugging Eq.~\eqref{eq:treeUVmodels} into~\eqref{eq:WCrun}, neglecting the small off-diagonal terms $[\hat{C}_i]_{1,2}$, we get the five scenarios outlined in Table~\ref{tab:UVtoWETscenarios}.
\begin{table}[]
    \centering
    \begin{tabular}{|c|c|c|c|c|c|}
    \hline
    & I & II & III & IV & V \\
    \hline &&&&&\\[-4mm]
        $\;C_{V_R}\;$ & $\;0.689 [\hat{C}_{V_R}]_3\;$ & 0 & 0& 0& 0\\[1mm]
        $\;C_S\;$ & 0 & $\;-0.57 ([\hat{C}_{S_L}]_3 + [\hat{C}_{S_R}]_3)\;$  & $\;2.34 [\hat{C}_T]_3\;$  & $\;-2.18[\hat{C}_T]_3\;$ & $\;-0.57[\hat{C}_{S_R}]_3\;$\\[1mm]
        $\;C_P\;$ & 0 &$\;0.57 ([\hat{C}_{S_L}]_3 - [\hat{C}_{S_R}]_3)\;$  & $\;-2.34 [\hat{C}_T]_3\;$  & $\;2.18[\hat{C}_T]_3\;$ & $\;-0.57[\hat{C}_{S_R}]_3\;$\\[1mm]
        $\;C_T\;$ & 0 &$\;0.003 [\hat{C}_{S_L}]_3\;$  & $\;-0.65 [\hat{C}_T]_3\;$  & $\;-0.63[\hat{C}_T]_3\;$ & $\;0\;$\\[1mm]
        \hline
    \end{tabular}
    \caption{Values for the WET Wilson coefficients at the low-scale $\mu = 5~\text{GeV}$ as a function of the SMEFT Wilson coefficients at 1~TeV in the five scenarios of~\eqref{eq:treeUVmodels}.}
    \label{tab:UVtoWETscenarios}
\end{table}
Hence we can parametrize the five scenarios in terms of our fit parameters as
\begin{align}
\label{eq:scen}
    (\text{I}) &\to \{[a_R], [\cVR]\}\,,\nonumber\\
    (\text{II}) &\to \{[a_S], [a_P], a_T = 0.0028\sqrt{a_S^2+a_P^2-2a_S a_P c(\delta_{SP})},[c(\delta_{SP})]\}\,,\nonumber\\
    (\text{III}) &\to \{a_S=3.6a_T, a_P=3.6a_T, [a_T], \cS=-1, \cP=1\}\,,\nonumber\\
    (\text{IV}) &\to \{a_S=3.5a_T, a_P=3.5a_T, [a_T], \cS=1, \cP=-1\}\,,\nonumber\\
    (\text{V}) &\to \{[a_S], a_P=a_S\} \,,
\end{align}
where we enclosed in square brackets the free parameters and we defined $\cos(\delta_{SP}) = \cos(\delta_{ST} -\delta_{PT})$.
All parameters not listed are set to zero.
We notice that in (I) and (II) we have a bidimensional and three-dimensional parameter space, respectively, while in the other cases it is always one-dimensional.

In order to compare the constraining power of our fit with that of fits to exclusive $B$ decays, we also explored two additional scenarios:
scenario VI, with only $\text{Re}[\tilde{C}_{V_R}]\neq 0$, and scenario VII involving  
$a_T \neq 0$ only.\footnote{This is completely equivalent to setting  $\text{Re}[\tilde{C}_T]\neq 0$ only, because one is sensitive to  $|\tilde{C}_T|^2 = a_T^2$ exclusively.}

\section{Numerical analysis}
\label{sec:numanal}
We now turn to the study of the numerical impact of NP in $\bar{B} \to X_c \ell \nub$ decays.

\subsection{Observables}
\label{sec:observables}
From the triple differential decay rate we build the first few moments of the kinematic distributions, as the OPE prediction is valid only at the level of integrated quantities.
We consider normalized moments in the lepton energy $E_\ell$, the hadronic invariant mass $m_X^2$ and the dilepton invariant mass $q^2$:
\begin{align}
\langle E_\ell^n \rangle_{E_\ell>{E_\ell}_{\rm cut}} &= \frac{\int_{{E_\ell}_{\rm cut}}^{{E_\ell}_{\rm max}} dE_\ell E_\ell^n \frac{d\Gamma}{dE_\ell}}{\int_{{E_\ell}_{\rm cut}}^{{E_\ell}_{\rm max}} dE_\ell \frac{d\Gamma}{dE_\ell}}\,,\nonumber\\
\langle m_X^{2n}\rangle_{E_\ell>{E_\ell}_{\rm cut}} &= \frac{\int_{{E_\ell}_{\rm cut}}^{{E_\ell}_{\rm max}} dE_\ell \int d u\, dq^2 m_X^{2n} \frac{d^3 \Gamma}{dE_\ell du dq^2}}{\int_{{E_\ell}_{\rm cut}}^{{E_\ell}_{\rm max}} dE_\ell \frac{d \Gamma}{dE_\ell}}\,,\nonumber\\
\langle q^{2n} \rangle_{q^2 > q^2_{\rm cut}} &= \frac{\int_{q^2_{\rm cut}}^{q^2_{\rm max}} dq^2 q^{2n} \frac{d\Gamma}{dq^2}}{\int_{q^2_{\rm cut}}^{q^2_{\rm max}} dq^2 \frac{d\Gamma}{dq^2}}\,,
\end{align}
where the hadronic invariant mass is given by
\begin{equation}
    m_X^2 = (p_B-q)^2 = \bar{\Lambda}^2 + m_b \bar{\Lambda}(1+\rho) + m_c^2 + (m_b + \bar{\Lambda})m_b \hat{u} - m_b \hat{q}^2\bar{\Lambda}\,,
\end{equation}
with $\bar{\Lambda} = m_B - m_b$. We will not expand our results in $\bar{\Lambda}$.
The kinematic upper limits for integration are ${E_\ell}_{\rm max} = (m_b^2 - m_c^2)/(2m_b)$ and $q^2_{\rm max} = (m_b - m_c)^2$. Lepton-energy and hadronic-invariant-mass moments are computed with a lower cut on the lepton energy ${E_\ell}_{\rm cut}$, while $q^2$ moments are computed with a lower cut $q^2_{\rm cut}$ on $q^2$.
We define therefore the first moments of interest as
\begin{equation}
    L_1({E_\ell}_{\rm cut}) \equiv \langle E_\ell \rangle_{E_\ell>{E_\ell}_{\rm cut}}\,, \qquad H_1({E_\ell}_{\rm cut}) \equiv \langle m_X^{2}\rangle_{E_\ell>{E_\ell}_{\rm cut}}\,, \qquad Q_1(q^2_{\rm cut}) \equiv \langle q^2 \rangle_{q^2 > q^2_{\rm cut}}\,.
\end{equation}
In order to reduce correlations between higher moments and the first one we consider central moments
\begin{align}
    L_{2,3}({E_\ell}_{\rm cut}) &= \langle (E_\ell - L_1({E_\ell}_{\rm cut}))^{2,3}\rangle_{E_\ell > {E_\ell}_{\rm cut}}\,,\nonumber\\
    H_{2,3}({E_\ell}_{\rm cut}) &= \langle (m_X^2 - H_1({E_\ell}_{\rm cut}))^{2,3}\rangle_{E_\ell > {E_\ell}_{\rm cut}}\,,\nonumber\\
    Q_{2,3}(q^2_{\rm cut}) &= \langle (q^2 - Q_1(q^2_{\rm cut}))^{2,3}\rangle_{q^2 > q^2_{\rm cut}}\,.
\end{align}
Furthermore we consider 
\begin{equation}
    R^*({E_\ell}_{\rm cut}) = \frac{\int_{{E_\ell}_{\rm cut}}^{{E_\ell}_{\rm max}} dE_\ell \frac{d\Gamma}{dE_\ell}}{\int_{0}^{{E_\ell}_{\rm max}} dE_\ell \frac{d\Gamma}{dE_\ell}}\,,
\end{equation}
such that the branching ratio with a lower cut on the lepton energy $\mathcal{B}_{\rm cut}({E_\ell}_{\rm cut})$ measured by BaBar~\cite{BaBar:2004bij,BaBar:2009zpz} and by Belle \cite{Belle:2006kgy} can be linked to the full branching ratio $\mathcal{B}_{\rm cut}(0) = \mathcal{B}(\bar{B}\to X_c \ell \nub)$
\begin{equation}
    \mathcal{B}_{\rm cut}({E_\ell}_{\rm cut}) = \mathcal{B}(\bar{B}\to X_c \ell \nub) R^*({E_\ell}_{\rm cut})\,,
\end{equation}
allowing for the extraction of $|\tilde{V}_{cb}|$ in the global fit~\cite{Gambino:2013rza}.
There are hence 10 experimental observables, each measured at some values for the cut (${E_\ell}_{\rm cut}$ or $q^2_{\rm cut}$) which we will compare to our theoretical expressions.
The moments in the lepton energy and hadronic invariant mass were measured by BaBar~\cite{BaBar:2004bij,BaBar:2009zpz}\footnote{The BaBar data for leptonic moments and $\mathcal{B}_{\rm cut}({E_\ell}_{\rm cut})$ are corrected for the complete $\mathcal{O}(\alpha_{\rm em})$ effects calculated in Ref.~\cite{Bigi:2023cbv}, see Section 3.2 of Ref.~\cite{Finauri:2023kte}.}, CLEO~\cite{CLEO:2004bqt}, DELPHI~\cite{DELPHI:2005mot}, CDF~\cite{CDF:2005xlh} and Belle~\cite{Belle:2006jtu,Belle:2006kgy}.
Measurements of the $q^2$-moments, suggested in Ref.~\cite{Fael:2018vsp}, have been performed by Belle~\cite{Belle:2021idw} and Belle II~\cite{Belle-II:2022evt}.
The data included in the fit are shown in Figs.~\ref{fig:plotObs} and~\ref{fig:plotR}. Additionally, 
we adopt the constraints on the $b$ and $c$ quark masses and on  $\mu_G^2$ and $\rho_{LS}^3$ 
employed in Ref.~\cite{Finauri:2023kte}.

The theoretical predictions are first computed in the on-shell scheme for the quark masses and the HQE parameters ($\mu^2_\pi$, $\mu^2_G$, $\rho^3_D$, $\rho^3_{LS}$). 
The expressions for $R^*$, the normalized and central moments are then expanded in $\as$, $\LamQCD/m_b$ and in the absolute values of the NP Wilson coefficients $a_i$  which are assumed to be small. We include terms up to $\mathcal{O}(a_i^2)$ or $\mathcal{O}(a_i a_j)$.
Afterwards we perform a change of scheme converting $m_b$ and the HQE parameters to the kinetic scheme with cut-off scale $\mu_k = 1~\text{GeV}$~\cite{Bigi:1996si,Fael:2020iea}, while the charm mass is converted into the $\overline{\text{MS}}$ scheme with associated scale $\mu_c = 2~\text{GeV}$.
After the change of scheme the expressions are re-expanded in $\as$, $\LamQCD/m_b$ and the $a_i$, but keeping all powers of the ratio $\mu_k/m_b$ which we count as an $\mathcal{O}(1)$ parameter.
Therefore a generic observable $\xi = \{R^*, L_n, H_n,Q_n, A_{FB}\}$ can be written as
\begin{equation}
	\begin{split}
    \label{eq:genmomxi}
		\xi & = \xi_{\textrm{SM}} + a_{R} \cVR \xi_{LR} + a_{R}^2 \bigl(\xi_{RR}+\cVR^2 \,\xi_{LR^2}\bigr) + a_{S}^2 \,\xi_{SS} + a_{P}^2 \,\xi_{PP} \\
		&+ a_{T}^2\, \xi_{TT} + a_S a_T  \cS\, \xi_{ST} + a_P a_T \cP\, \xi_{PT}\,,
	\end{split}
\end{equation}
where $\xi_{\rm SM}$ is the SM prediction including corrections up to $\mathcal{O}(\as^2)$~\cite{Melnikov:2008qs,Pak:2008cp,Fael:2024gyw}, $\mathcal{O}(\as/m_b^2)$~\cite{Alberti:2012dn,Alberti:2013kxa} and $\mathcal{O}(1/m_b^3)$~\cite{Blok:1993va,Manohar:1993qn,Gremm:1996df}, except for the $q^2$ moments where also the $\mathcal{O}(\as/m_b^3)$ corrections~\cite{Mannel:2021zzr} are available.
The inclusion of the full $\mathcal{O}(\as^2)$ corrections to the $q^2$ moments, computed recently in Ref.~\cite{Fael:2024gyw}, updates the SM theoretical predictions presented in Ref.~\cite{Finauri:2023kte}, where only the $\mathcal{O}(\as^2 \beta_0)$ terms (the so-called BLM approximation) were considered. In the case of $A_{FB}$ we do not include $\mathcal{O}(\alpha_s)$ corrections, as this observable does not enter the fit.
The NP coefficients $\xi_{XY}$ are computed up to order $\mathcal{O}(1/m_b^3)$ at tree level in QCD, except for $\xi_{LR}$, $\xi_{RR}$ and $\xi_{LR^2}$, where also $\mathcal{O}(\as)$ one-loop QCD corrections to the leading power term are included, since they come from the only operator which has non-vanishing interference with the SM.
In the case of the total width $\Gamma$, which will be used for the extraction of $|\tilde{V}_{cb}|$, we have a similar expression
\begin{equation}
\label{eq:gammatildes}
	   \Gamma = \Gamma_0 \biggl(\zeta_{\textrm{SM}} + a_R \cVR \zeta_{LR}+ a_{R}^2 \zeta_{RR} + a_{S}^2 \zeta_{SS} + a_{P}^2 \zeta_{PP} + a_{T}^2 \zeta_{TT}\biggr)\,,
\end{equation}
where we have defined
\begin{equation}
\label{eq:G0}
    \Gamma_0 = |\tilde{V}_{cb}|^2 A_{\rm ew} \frac{G_F^2 m_b^5}{192\pi^3}\,,
\end{equation}
with the standard universal electroweak correction factor $A_{\rm ew} = 1 + \frac{\alpha_{\rm em}}{\pi}\ln \frac{m_Z^2}{m_b^2} = 1.014$ coming from $|C_{V_L}|^2$.
As noticed below Eq.~\eqref{eq:d2Gnoml}, the interference terms between $C_S$, $C_P$ and $C_T$ vanish in $\Gamma$, while the term $a_{R}^2 \cVR^2$ is missing due to the fact that no denominator has been re-expanded in this expression.
In the SM prediction of the decay rate we include perturbative corrections up to $\mathcal{O}(\as^3)$~\cite{Fael:2020tow} at leading power, and $\mathcal{O}(\as/m_b^3)$~\cite{Mannel:2019qel} for the power corrections.
The NLO non-logarithmic QED and electroweak corrections are also included~\cite{Bigi:2023cbv}, on top of the universal electroweak correction $A_{\rm ew}$ already introduced in Eq.~\eqref{eq:G0}.
The NP part is computed at the same level of accuracy as for the moments~\eqref{eq:genmomxi}.

We present numerical results for the observables $L_n(1~\text{GeV})$, $H_n(1~\text{GeV})$, $Q_n(4~\text{GeV}^2)$, $R^*(1~\text{GeV})$, $A_{FB}(4~\text{GeV}^2)$  and $\Gamma$ in Tables~\ref{tab:El_moments}, \ref{tab:MX_moments}, \ref{tab:q2_moments}, \ref{tab:Rstar}, \ref{tab:AFBtab} and \ref{tab:Gammatot}, respectively, gathered in Appendix \ref{app:num}. The tables are obtained with the input values listed in Table~\ref{tab:numeric input}, and with $\as^{(n_f=4)}(\mu_s = m_b) = 0.216$.
As is well-known, the central moments tend in general to be more sensitive to higher-order corrections, in particular power corrections.

It is worth noticing that the coefficient $\xi_{SS}$ at leading power receives a suppression of order $\sqrt{\rho}$, which is not present for the power-correction terms.
This arises from a partial cancellation between the numerator and denominator in normalized quantities, and in the case of the leptonic moments the cancellation  is  due to the similarity of the lepton energy spectra generated by the scalar and the SM operators.

Furthermore, for $m_\ell=0$, the vector-vector contributions to $\Gamma$ only come from $w_1$ and $w_2$, as can be seen from Eq.~\eqref{eq:d2Gnoml}.
The transformation properties of $W_{\Gamma,\Gamma'}$ under parity imply the relation
\begin{equation}
    w_{1,2} = w_{1,2}\big|_{C_{V_L} \leftrightarrow C_{V_R}}\,,
\end{equation}
as in the symmetric terms of the hadronic tensor only an even number of $\gamma^5$ matrices gives a non-vanishing contribution, making the sign in the chiral projector irrelevant.
We therefore have $\zeta_{RR} = \zeta_{\rm SM}$ which holds to all orders in the perturbative and power expansions.
In a similar way, the $a_R^2$ terms cancel in normalized $q^2$ moments between numerator and denominator, as $a_R^2$ contributes as a rescaling of the SM term by $(1+a_R^2)$, resulting in $\xi_{RR}=0$.
Finally, we note that the coefficient $\zeta_{TT}$ defined in Eq.~\eqref{eq:gammatildes} is one order of magnitude larger than all the other coefficients. This will play an important role in the fit performed in the next section.
The $\xi_{TT}$ coefficients in the moments exhibit a similar pattern, although the effect is generally reduced due to a partial cancellation between the numerator and denominator.
The origin of the enhancement is a combination of purely combinatoric factors, given the higher number of terms in the contractions of Lorentz indices in the tensor-tensor contribution, and an enhancement of the tensor contribution close to $q^2=0$, similar to the one observed in $\bar B\to D^*\ell\bar\nu$ decays \cite{Jung:2018lfu}.

\subsection{Global fit results}
\label{sec:globfit}
The fit is performed with the method of least squares, minimizing the $\chi^2$ function.
The theoretical uncertainties and correlations are handled in the same way as in Refs.~\cite{Gambino:2013rza,Finauri:2023kte}.
The $\chi^2$ is approximately linear in the SM fit parameters $\{m_b,m_c,\mu^2_\pi,\mu^2_G,\rho^3_D,\rho^3_{LS},\text{BR}_{c\ell\bar{\nu}}\}$, when expanding around their central values.
When adding the NP parameters $\{a_i\}$ and $\{\cVR,\cS,\cP\}$, the $\chi^2$ is in general non-linear, and confidence level regions can be obtained by 
profiling the $\Delta \chi^2 = \chi^2 - \chi^2_{\rm min}$ over the parameters of interest. 
Due to Wilks' theorem the $\Delta\chi^2$ follows a $\chi^2$ distribution with number of degrees of freedom equal to the dimension of the scanned parameter subspace.
The value of $|\tilde{V}_{cb}|$ is then extracted from the fit results and the theoretical expression~\eqref{eq:gammatildes} for $\Gamma$.
Due to the truncation of the perturbative and power expansions in $\Gamma$, we include an additional theoretical uncertainty on $|\tilde{V}_{cb}|$, $\sigma_\Gamma = 0.00025$, following Ref.~\cite{Bordone:2021oof}.
Our choices for the reference values for the kinetic-scheme scale and the scales at which $m_c$ and $\alpha_s$ are evaluated follow  Ref.~\cite{Finauri:2023kte}, namely $\mu_k=1~\text{GeV}$, $\mu_c = 2~\text{GeV}$ and $\mu_s = m_b/2$ for the scale of $\alpha_s^{(4)}$.
As a starting point we  perform a SM  fit, which results in $\chi^2 / \text{d.o.f.} = 42.58/74$, which one might consider indicative of our conservative theory uncertainties, see the discussion in Ref.~\cite{Gambino:2013rza}.
In this case, the fit becomes approximately linear, allowing analytical minimization of the $\chi^2$.
The result of this SM fit, including correlations among the fit parameters, is shown in Table~\ref{tab:fitSMres}. The value of $|V_{cb}|$
 is 0.8\% lower than in Ref.~\cite{Finauri:2023kte} because of the inclusion of the complete $\mathcal{O}(\alpha_s^2)$ corrections to the $q^2$ moments\footnote{Recently, the results of Ref.~\cite{Fael:2024gyw} have been confirmed in Ref.~\cite{Czaja:2024tcv}, where triple-charm contributions were also computed and found to be negligible.}~\cite{Fael:2024gyw} and because of the inclusion of the total $\mathcal{O}(\alpha_{\rm em})$ contribution to the decay width~\cite{Bigi:2023cbv}.
\begin{table}
    \centering
\begin{tabular}{cccccccc}
 \hline
&&&&&&&\\[-4.3mm]
$m_b\,[\text{GeV}]~$ & $m_c\,[\text{GeV}]~$ & $\mu_\pi^2\,[\text{GeV}^2]~$ & $\mu_G^2\,[\text{GeV}^2]~$ & $\rho^3_D\,[\text{GeV}^3]~$ & $\rho^3_{LS}\,[\text{GeV}^3]~$ & $\rm{BR}_{c\ell\bar{\nu}}\,[\%]~$ & $10^3|{V}_{cb}|$ \\[1.2mm]
\hline
&&&&&&&\\[-4.5mm]4.574 & 1.090 & 0.435 & 0.278 & 0.164 & -0.090 & 10.62 & 41.64 \\
0.012 & 0.010 & 0.040 & 0.048 & 0.018 & 0.089 & 0.15 & 0.47\\
\hline &&&&&&&\\[-4.3mm]
 1 & 0.390 & -0.229 & 0.560 & -0.022 & -0.181 & -0.064 & -0.421 \\
  & 1 & 0.015 & -0.238 & -0.028 & 0.084 & 0.034 & 0.071 \\
  &  & 1 & -0.097 & 0.535 & 0.266 & 0.144 & 0.346 \\
  &  &  & 1 & -0.261 & 0.004 & 0.001 & -0.271 \\
  &  &  &  & 1 & -0.014 & 0.025 & 0.172 \\
  &  &  &  &  & 1 & -0.010 & 0.056 \\
  &  &  &  &  &  & 1 & 0.694 \\
  &  &  &  &  &  &  & 1 \\
\hline
\end{tabular}
    \caption{Results for the global fit with SM only. Central values on the first line, uncertainties on the second line and correlation matrix below.}
    \label{tab:fitSMres}
\end{table}

We now add the NP degrees of freedom to the fit, numerically minimizing the $\chi^2$ for each NP scenario of~\eqref{eq:scen}, as well as for the full set of NP parameters.
The resulting set of minimum $\chi^2$ is
\begin{align}
\label{eq:chimins}
    \chi^2_{\rm min}/\text{d.o.f.}|_{\text{(I)}} &= 40.06/72\,, \qquad \chi^2_{\rm min}/\text{d.o.f.}|_{\text{(II)}} = 42.52/71\,, \nonumber\\
    \chi^2_{\rm min}/\text{d.o.f.}|_{\text{(III)}} &= 42.58/73\,, \qquad \chi^2_{\rm min}/\text{d.o.f.}|_{\text{(IV)}} = 41.12/73\,, \nonumber\\
    \chi^2_{\rm min}/\text{d.o.f.}|_{\text{(V)}} &= 42.52/73\,, \qquad \chi^2_{\rm min}/\text{d.o.f.}|_{\text{(NP)}} = 36.04/67\,,
\end{align}
with the scenarios I-V corresponding to single-mediator models with a vector-like quark doublet, a second Higgs doublet, and three different leptoquark models, all specified in Eqs.~\eqref{eq:treeUVmodels} and \eqref{eq:scen}, while 'NP' refers to the global fit with all WET contributions present simultaneously.
We note that scenarios II, III and V do not improve the description of the data with respect to the pure SM fit, while scenarios I and IV slightly decrease the minimal $\chi^2$, albeit not significantly.
The same applies to the general NP case, where the $\chi^2$ decreases roughly by the number of degrees of freedom added to the fit.
In fact, the best fit point in the general NP case is compatible with the SM fit at the $0.7\sigma$ level, assuming a $\chi^2$ distribution with $\Delta\chi^2 = 6.54$ and 7 degrees of freedom.
Given the high level of agreement between the data and the SM predictions on the one side, and the at most moderate reduction in $\chi^2$ when including NP contributions on the other, we do not interpret these results as indicative of the presence of BSM physics.

\begin{figure}
    \centering
    \includegraphics[width=0.99\linewidth]{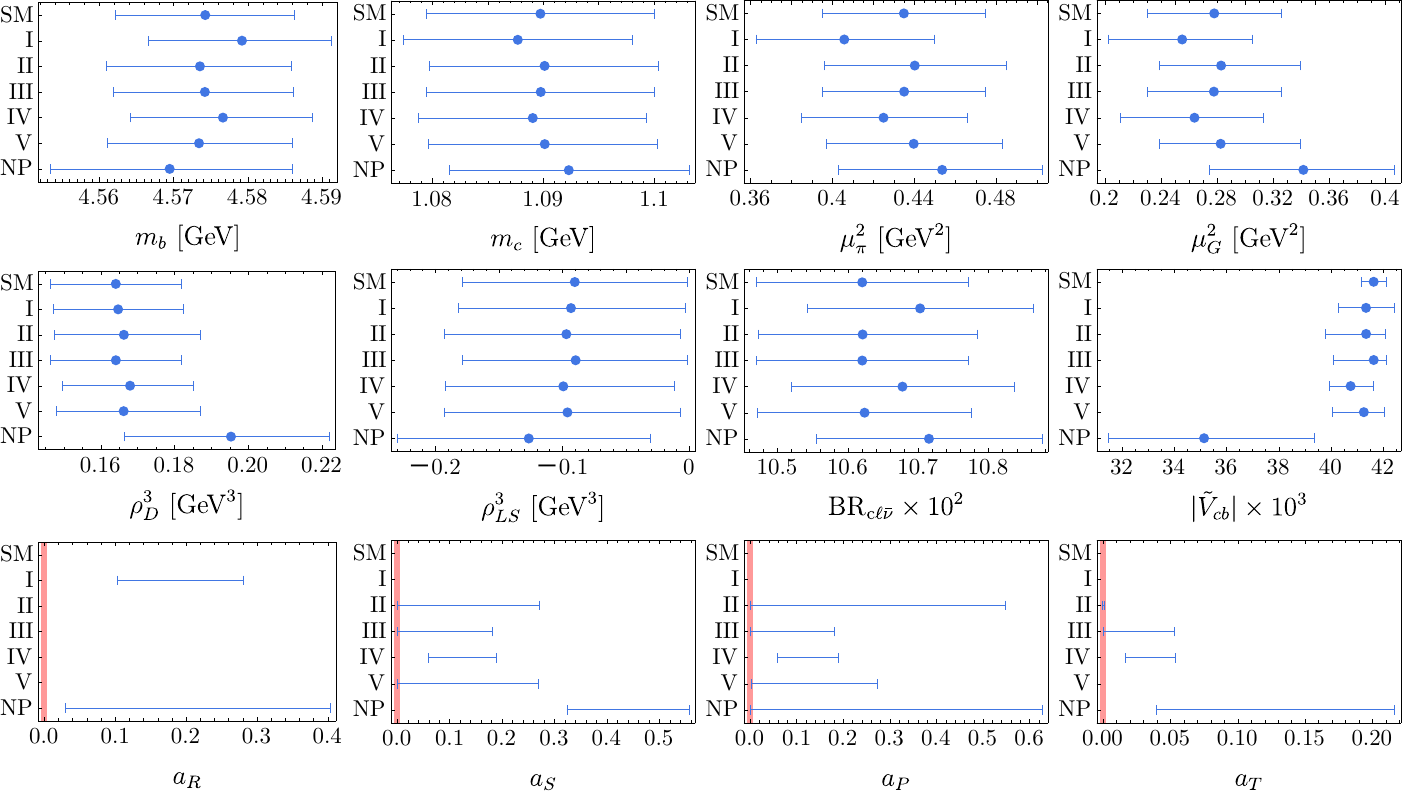}
     \vspace{-0.4cm}
    \caption{Fit results in the SM, the five scenarios of~\eqref{eq:scen}, and the full NP case. The error bands correspond to $\Delta \chi^2 \leq 1$,
    the red line denotes the boundary condition. The central values for the NP parameters are not shown as their statistical distribution is not Gaussian.}
    \label{fig:scenPlot}
\end{figure}
The fit results assuming the SM alone, the five NP scenarios~\eqref{eq:scen} or the full set of NP operators are displayed in Fig.~\ref{fig:scenPlot} (results for $\cVR$, $\cS$, $\cP$ and $\cos(\delta_{SP})$ are not displayed as they do not provide much information).
We notice that every scenario gives a result consistent with the value of $|\tilde{V}_{cb}|$ preferred by the SM fit, in accordance with the moderate reduction of the $\chi^2$ observed above.
It is important to stress that the NP parameters do not follow a Gaussian distribution, and hence the quoted intervals do not allow to infer  the size of the intervals at higher $\Delta\chi^2$ values, which are in particular necessary to judge the overall tension with the SM. 
For this reason we provide in Appendix~\ref{sec:appplots} the 1D-profile plots for $\Delta \chi^2$ as a function of the NP parameters and $|\tilde{V}_{cb}|$.

Interpreting the maximum value for the $a_i$ of $\sim 0.6$ as a bound on the NP scale by setting\footnote{This assumes the high energy NP coupling parameter to be unity.} $\LamNP \simeq (4G_F |V_{cb}| a_i/\sqrt{2})^{-1/2}$, we obtain $\LamNP \gtrsim 1.1~\text{TeV}$, consistent with the typical scales associated to flavour anomalies, see for instance Refs.~\cite{Albrecht:2021tul,Capdevila:2023yhq}. This justifies our initial hypothesis of working in WET. 

\begin{figure}
    \centering
    \includegraphics[width=0.33\textwidth]{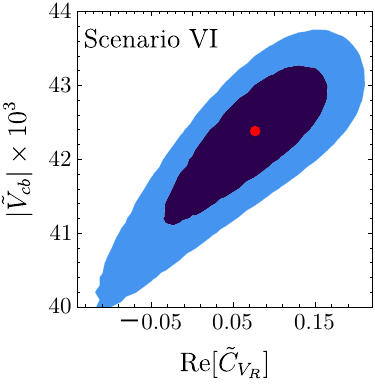}~
    \includegraphics[width=0.33\textwidth]{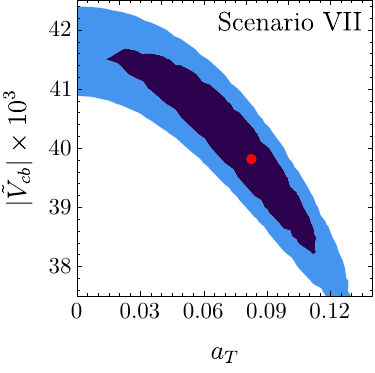}~
    \includegraphics[width=0.33\textwidth]{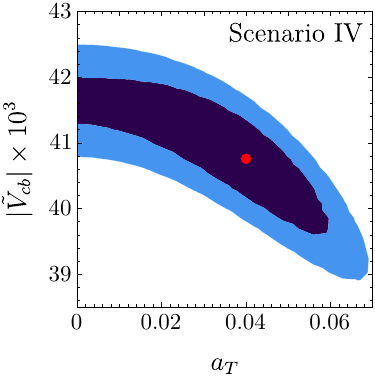}
     \vspace{-0.6cm}
    \caption{2D profile $\Delta \chi^2$ plot in the $\text{Re}[\tilde{C}_{V_R}]-|\tilde{V}_{cb}|$ (left) and $a_T-|\tilde{V}_{cb}|$ (right) planes. The red dot stands for the best fit point $\Delta \chi^2 =0$, while the dark blue and light blue regions correspond to $\Delta\chi^2 \leq 2.30$ and $\Delta\chi^2 \leq 5.99$, respectively.}
    \label{fig:plot2D_CT}
\end{figure}

Combined fits of exclusive and inclusive measurements have been performed in Refs.~\cite{Crivellin:2014zpa,Jung:2018lfu,Iguro:2020cpg} using the total inclusive decay rate computed at tree level without nonperturbative corrections.
In order to compare the constraining power of our fit with these results, we introduced scenarios VI and VII  in the previous section.
In Fig.~\ref{fig:plot2D_CT} we show 2D $\chi^2$ contour plots of the fits with the right-handed vector operator only (scenario VI, left panel), the tensor operator only (scenario VII, central panel), as well as scenario IV (right panel).
As expected, the constraints we obtain from a fit to the full set of kinematic moments are much stronger than those previously obtained from the decay rate alone.
As an effect, the allowed parameter space changes qualitatively: while previously the single constraint from the total rate resulted only in an extended band in this space, the new fit constrains the plotted parameters individually. 
The resulting constraints are complementary and competitive with the ones from exclusive modes \cite{Crivellin:2014zpa,Jung:2018lfu,Iguro:2020cpg,Fedele:2023ewe,Ray:2023xjn}.

The case of NP in $\textrm{Re}[\tilde{C}_{V_R}]$ deserves some additional comments. 
It was investigated in particular in Ref.~\cite{Feger:2010qc}, which stands out as the only available work in the literature performing a fit to inclusive $\bar{B} \to X_c \ell \nub$ moments including NP contributions.
Using constraints from the lepton-energy and hadronic-invariant-mass moments only, the authors found a blind direction in $\tilde{V}_{cb} - \textrm{Re}[\tilde{C}_{V_R}]$ (Fig.~3 of \cite{Feger:2010qc}), rendering the projection of the bound on $\textrm{Re}[\tilde{C}_{V_R}]$ uncompetitive. 
We agree with this observation, but remarkably the $q^2$ moments provide a strong constraint that does not suffer from this problem, and thereby the dominant bound on $\tilde{C}_{V_R}$.

In a generic scenario with a right-handed vector current, considering only the real part of $\tilde{C}_{V_R}$ may be motivated by the fact that the imaginary part of $\tilde{C}_{V_R}$ enters only quadratically in a small Wilson coefficient expansion of the observables. 
However, when we allow for an imaginary part of $\tilde{C}_{V_R}$, we find that the fit results change drastically, and 
that the best fit point of scenario I (i.e., allowing for a complex $C_{V_R}$) is complex ($\cVR \neq \pm1$ see Fig.~\ref{fig:1DI}).
Therefore, neglecting the imaginary part of $\tilde{C}_{V_R}$ is not a good approximation in general.

\begin{table}
    \centering
\begin{tabular}{ccccccc}
 \hline
&&&&&&\\[-4.3mm]
$m_b\,[\text{GeV}]~$ & $m_c\,[\text{GeV}]~$ & $\mu_\pi^2\,[\text{GeV}^2]~$ & $\mu_G^2\,[\text{GeV}^2]~$ & $\rho^3_D\,[\text{GeV}^3]~$ & $\rho^3_{LS}\,[\text{GeV}^3]~$ & $\rm{BR}_{c\ell\bar{\nu}}\,[\%]$ \\[1.2mm]
\hline
&&&&&&\\[-4.5mm]
4.569 & 1.092 & 0.454 & 0.342 & 0.195 & $-0.126$ & 10.72 \\[1.1mm]
$+0.017\atop{-0.016}$ & $+0.011 \atop -0.011$ & $+0.049 \atop -0.050$ & $+0.065 \atop -0.067$ & $+0.027 \atop -0.029$ & $+0.096 \atop -0.104$ & $+0.16\atop -0.16$\\[1.5mm]
\hline
&&&&&&\\[-4.8mm]
$a_{R}~$ & $a_S$ & $a_P$ & $a_T$ & $\cVR$ & $\cS$ & $\cP$ \\[1.mm]
\hline
&&&&&&\\[-4.5mm]
0.19 & 0.46 & 0.20 & 0.15 & 0.99 & $-0.94$ & 0.91 \\[1.1mm]
$+0.21\atop{-0.16}$ & $+0.10 \atop -0.14$ & $+0.43 \atop -0.20$ & $+0.07 \atop -0.11$ & $+0.01 \atop -0.81$ & $+0.92 \atop -0.06$ & $+0.09\atop -1.91$\\[1.5mm]
\hline
\end{tabular}
    \caption{Results for the global fit with full set of NP operators. Central values on the first line, $68.3\%$ confidence level intervals on the second line. The value of $|\tilde{V}_{cb}|$ extracted from the fit parameters is shown in~\eqref{eq:VcbNPfit}.}
    \label{tab:fitNPres}
\end{table}
We finally turn to the case of the global fit with all NP parameters present simultaneously.
The situation changes significantly in this case, as large cancellations between the $a_T^2$, $a_S a_T \cS$ and $a_P a_T \cP$ terms allow for markedly larger NP contributions.
This results in a much lower value for $|\tilde{V}_{cb}|$,
\begin{equation}
\label{eq:VcbNPfit}
    |\tilde{V}_{cb}| = \Bigl(35.1^{+4.2}_{-3.6}\Bigr) \cdot 10^{-3}\,,
\end{equation}
where the uncertainty is almost ten times larger than in the SM.
The results for the fit parameters are shown in Table~\ref{tab:fitNPres}, where the parameter intervals are obtained by profiling the $\chi^2$ in one dimension.

In Fig.~\ref{fig:NPfit} we present the results of the fit in the form of 2D $\chi^2$ profile plots for the absolute values $a_i$ and $|\tilde{V}_{cb}|$.
\begin{figure}
    \centering
    \includegraphics[width=\linewidth]{Img/profile2D_fullNP}
     \vspace{-0.5cm}
    \caption{2D contour plots for pairs of fit parameters in the global fit with $\Delta \chi^2 \leq 2.30$ (dark blue) and $\Delta\chi^2 \leq5.99$ (light blue)
    (plots for the phases are not shown for simplicity). The red dot indicates the best fit point.}
    \label{fig:NPfit}
\end{figure}
In particular, the last panel in the lowest row shows that there is a high degree of correlation between $a_{T}$ and $|\tilde{V}_{cb}|$.
Therefore the low value of $|\tilde{V}_{cb}|$ and the large value of $a_T$ are related: the large $a_T$ needs to be compensated by an anomalously low $|\tilde{V}_{cb}|$ value, because of the large coefficient $\zeta_{TT}$ in Table~\ref{tab:Gammatot}. 
Therefore, the 1D confidence level interval given in Eq.~\eqref{eq:VcbNPfit} should be taken with a pinch of salt,
and indeed, as $a_T$ approaches zero, the value of $|\tilde{V}_{cb}|$ gets closer and closer to the result of the SM fit.
Additional observables, like those from exclusive semileptonic $B$ decays or the forward-backward asymmetry in inclusive decays~\cite{Turczyk:2016kjf,Fael:2022wfc,Herren:2022spb} (see Eq.~\eqref{eq:AFB}), could potentially remove this degeneracy and improve significantly the constraining power of the fit.

\begin{figure}
    \centering
    \includegraphics[width=\linewidth]{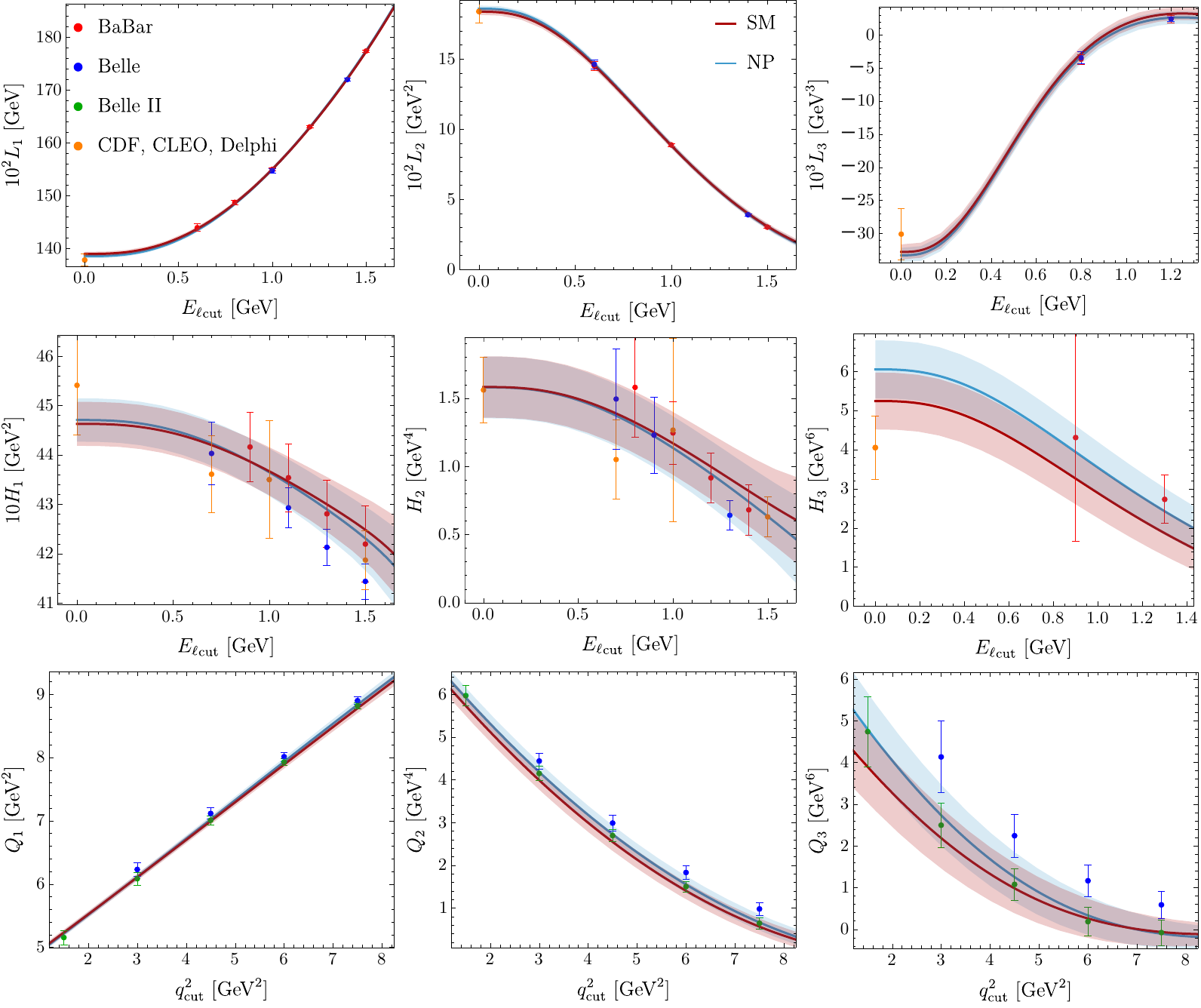}
     \vspace{-0.5cm}
    \caption{Observables included in the fit, with the theoretical predictions obtained from the SM (red) and full NP (blue) fit results. The shaded bands denote the theoretical uncertainties.}
    \label{fig:plotObs}
\end{figure}

\begin{figure}
    \centering
    \includegraphics[width=0.48\linewidth]{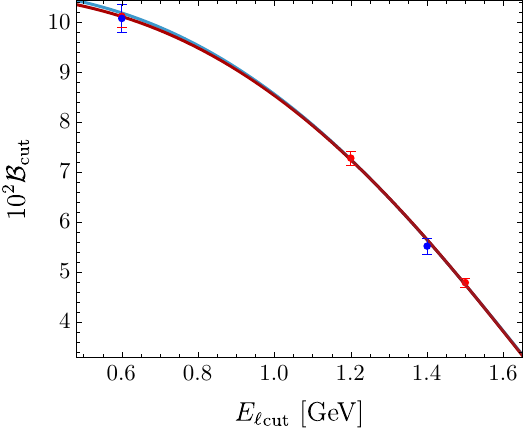}
     \vspace{-0.4cm}
    \caption{Theoretical predictions for $\mathcal{B}_{\rm cut}({E_\ell}_{\rm cut})$ compared to the experimental data from BaBar (red) and Belle (blue). The red line denotes the SM prediction while the blue one includes NP contributions.}
    \label{fig:plotR}
\end{figure}

The theoretical predictions resulting from the SM and global NP fits are shown in Figs.~\ref{fig:plotObs} and~\ref{fig:plotR} with error bands denoting the assigned theoretical uncertainties, which follow the same strategy as the one employed in Ref.~\cite{Finauri:2023kte}.
The plots also show the experimental measurements included in the global fit.
We observe only small shifts in most of the observables. An interesting aspect concerns the small tension between the results for the $q^2$ moments from Belle and Belle~II: while in the SM the constraints from lepton-energy and hadronic moments result in a clear preference of the fit for the moments as measured by Belle~II, in the global fit this preference is less pronounced. A resolution of this tension will therefore affect the results for BSM physics from inclusive decays.

\section{Summary and prospects}
\label{sec:conclusion}

We have presented a comprehensive study of BSM physics in inclusive semileptonic $B$ decays including all dimension-six NP operators in WET. 
We have computed the BSM contributions up to $\mathcal{O}(\LamQCD^3/m_b^3)$ in the HQE, reproducing many of the results given in the literature, see Table \ref{tab:NPchecks}. 
We have also reproduced the $\mathcal{O}(\alpha_s)$ corrections for the right-handed vector current in Refs.~\cite{Dassinger:2007pj,Alberti:2015qmj}. 
Explicit analytic expressions for the structure functions are provided in the ancillary files.
In addition to the generic WET case, we have considered five single-mediator scenarios, matched first to SMEFT then to WET.

We have performed the first global fit to the full set of observables in inclusive semileptonic $B$ decays using the entire WET basis of five dimension-six operators.
We do not observe any significant indication of BSM contributions to the investigated observables.
The analysis provides qualitatively different constraints from those observed before, and results in complementary and competitive constraints with respect to exclusive
measurements for the pertinent WET coefficients. 
The analysis yields a lower limit for the effective energy scale of heavy NP of the order of $\mathcal{O}(1~\text{TeV})$, comparable with other BSM studies in $B$ physics.
The constraining power of the available inclusive data with respect to individual parameters is significantly reduced once all of the NP Wilson coefficients are kept complex and free to float (seven independent real parameters).
Indeed, a flat direction arises due to the interference between the tensor and (pseudo)scalar operators, leading to an anomalously low central value of $|\tilde{V}_{cb}| = \bigl(35.1^{+4.2}_{-3.6}\bigr) \cdot 10^{-3}$. 
We have verified that a future measurement of the forward-backward asymmetry at Belle II could eliminate that flat direction.
Importantly, we observe that the values of the HQE parameters selected by the fit are rather stable under the inclusion of the BSM contributions.
In specific single-mediator models inclusive observables alone provide already competitive bounds, with best-fit values of $|\tilde{V}_{cb}|$ close to the SM fit for all of them: 
if it were not for the flat direction discussed above, our results show that NP cannot lower the inclusive value of $|\tilde{V}_{cb}|$ significantly.
A significant reduction of $|\tilde{V}_{cb}|$ is only observed in the presence of several NP operators, implies large cancellations between different contributions, and seems even in this case to be related to the aforementioned flat direction in the fit.

A global fit to inclusive and exclusive measurements is the next natural step to further constrain the parameter space, as both classes of processes are sensitive to the same WET NP operators. 
Furthermore, the treatment of hadronic matrix elements is completely independent in the two classes, and a combined fit could hence improve the separation between nonperturbative inputs and NP parameters.

The NP sensitivity of the global fit would also be enhanced if one could rely on an independent method to determine the nonperturbative HQE parameters or the full hadronic matrix elements of the WET operators. 
Lattice QCD calculations have this potential, and there are two ways in which they can contribute. 
The first consists in a determination of the HQE parameters from a precise calculation of the pseudoscalar- and vector-meson masses as functions of the heavy quark mass~\cite{Gambino:2017vkx,Brambilla:2017hcq}.
Here a limitation is related to the presence of additional nonperturbative parameters, the matrix elements of non-local operators. 
The second and more promising way in which lattice QCD can contribute is through the direct calculation on the lattice of the inclusive decay rate and of inclusive quantities such as the leptonic moments \cite{Gambino:2019sif, Gambino:2022dvu}. 
This method can be applied both in the SM and for NP, has been recently validated on $D_s$ data \cite{DeSantis:2025yfm,Kellermann:2025pzt}, 
and will soon be tested in the case of $B_s$ and $B$ meson decays, which might in time enhance the sensitivity of the global fit to NP degrees of freedom.

Including this kind of improvements into our analysis together with the expected high-precision results for inclusive decays from the Belle~II experiment \cite{Belle-II:2018jsg} will allow for even more stringent tests of BSM physics in $b\to c\ell\bar\nu$ transitions in the future.

\subsubsection*{Acknowledgements}
We would like to thank Fulvia De Fazio, Pietro Colangelo and Francesco Loparco for useful correspondence about Ref.~\cite{Colangelo:2020vhu}.
We also thank Markus Prim, Matteo Fael and Soumitra Nandi for helpful communications.
Some of the calculations performed in this work were done with the help of \texttt{FeynCalc}~\cite{Shtabovenko:2020gxv,Shtabovenko:2023idz}.
This work is supported 
by the Italian Ministry of University and Research (MUR) and the European Union (EU) – Next Generation EU, Mission 4, Component 1, PRIN 2022, grant 2022N4W8WR, CUP D53D23002830006.

\appendix
\section{Numerical results}
In this section we present numerical results for the coefficients introduced in Eqs.\ \eqref{eq:genmomxi} and \eqref{eq:gammatildes}, encoding the effect of NP in the prediction of the kinematic moments, $R^*$, $A_{FB}$ and $\Gamma$. The input parameters used, together with $\as^{(n_f=4)}(\mu_s = m_b) = 0.216$ and $\alpha_{\rm em} = 0.00757$, are listed in Table \ref{tab:numeric input}.
\setlength{\tabcolsep}{12pt} 
\label{app:num}
\begin{table}[H]
    \centering
    \begin{tabular}{|c|c|}
    \hline
       $m_b^{\textrm{kin}}(\rm 1\, GeV)$ & $4.573$ \textrm{GeV} \\ \hline
       $\overline{m}_c(\rm 2\,GeV)$ & $1.090$ \textrm{GeV} \\ \hline
        $\mu_\pi^2(\rm 1\, GeV)$ & $0.454$ $\textrm{GeV}^2$\\ \hline
        $\mu_G^2(\rm 1\, GeV)$ & $0.288$ $\textrm{GeV}^2$\\ \hline
        $\rho_{D}^3(\rm 1\, GeV)$ & $0.176$ $\textrm{GeV}^3$ \\ \hline
        $\rho_{LS}^3(\rm 1\, GeV)$ & $-0.113$ $\textrm{GeV}^3$ \\ \hline
    \end{tabular}
    \caption{Numerical inputs for Tables~\ref{tab:El_moments} to~\ref{tab:Gammatot}.}
    \label{tab:numeric input}
\end{table}

\setlength{\tabcolsep}{1pt} 
\begin{table}[H]
    \centering
    \begin{adjustbox}{max width=\textwidth}
    \begin{tabular}{|c|rll|rll|rll|}
    \hline
 & \multicolumn{3}{c|}{$L_1 ~[10^{-2} \ \textrm{GeV}]$} & \multicolumn{3}{c|}{$L_2 ~ [10^{-2} \ \textrm{GeV}^2]$} & \multicolumn{3}{c|}{$L_3~[10^{-3} \ \textrm{GeV}^3]$} \\ \hline
\multirow{2}{*}{$~\xi_{SM}~$} & $157.276$&$-1.676_{\textrm{pow}}$&$-0.312_{\alpha_s} $ & $8.726$&$+0.279_{\textrm{pow}}$&$-0.034_{\alpha_s} $ & $-3.096$&$+3.201_{\textrm{pow}}$&$+0.986_{\alpha_s} $ \\ 
&& $-0.463_{\frac{\alpha_s}{m_b^2}} $ & $+0.084_{\alpha_s^2}$ & & $-0.192_{\frac{\alpha_s}{m_b^2}} $ & $+0.057_{\alpha_s^2}$ & & $-0.364_{\frac{\alpha_s}{m_b^2}} $ & $+0.672_{\alpha_s^2}$ \\ \hline
$~\xi_{RR}~$ & $-9.760$&$+1.114_{\textrm{pow}}$&$+0.208_{\alpha_s}$ & $0.004$&$-0.287_{\textrm{pow}}$&$+0.021_{\alpha_s}$ & $9.004$&$-1.541_{\textrm{pow}}$&$-0.431_{\alpha_s}$ \\ \hline
$~\xi_{LR}~$ & $-0.368$&$+0.606_{\textrm{pow}}$&$-0.197_{\alpha_s}$ & $0.279$&$+0.105_{\textrm{pow}}$&$-0.013_{\alpha_s}$ & $0.576$&$+0.235_{\textrm{pow}}$&$+0.047_{\alpha_s}$ \\ \hline
$~\xi_{LR^2}~$ & $-0.247$&$+0.427_{\textrm{pow}}$&$-0.128_{\alpha_s}$ & $0.186$&$+0.060_{\textrm{pow}}$&$-0.013_{\alpha_s}$ & $0.418$&$+0.087_{\textrm{pow}}$&$+0.040_{\alpha_s}$ \\ \hline
$~\xi_{TT}~$ & $-78.076$&$+9.286_{\textrm{pow}}$& & $0.033$&$-1.071_{\textrm{pow}}$& & $72.030$&$-6.863_{\textrm{pow}}$& \\ \hline
$~\xi_{SS}~$ & $0.184$&$+3.006_{\textrm{pow}}$& & $-0.139$&$+0.471_{\textrm{pow}}$& & $-0.288$&$-2.420_{\textrm{pow}}$& \\ \hline
$~\xi_{PP}~$ & $-0.184$&$+0.191_{\textrm{pow}}$& & $0.139$&$+0.030_{\textrm{pow}}$& & $0.288$&$-0.169_{\textrm{pow}}$& \\ \hline
$~\xi_{ST}~$ & $-19.519$&$-2.147_{\textrm{pow}}$& & $0.008$&$-1.988_{\textrm{pow}}$& & $18.007$&$-3.218_{\textrm{pow}}$& \\ \hline
$~\xi_{PT}~$ & $19.519$&$-0.025_{\textrm{pow}}$& & $-0.008$&$+0.776_{\textrm{pow}}$& & $-18.007$&$+0.502_{\textrm{pow}}$& \\ \hline
\end{tabular}
    \end{adjustbox}
    \vspace{-0.3cm}
    \caption{Lepton energy moments with ${E_\ell}_{\rm cut} = 1~\textrm{GeV}$.}
    \label{tab:El_moments}
\end{table}

\begin{table}[H]
    \centering
    \begin{adjustbox}{max width=\textwidth}
    \begin{tabular}{|c|rll|rll|rll|}
    \hline
 & \multicolumn{3}{c|}{$H_1 ~ [10^{-1} \ \textrm{GeV}^2]$} & \multicolumn{3}{c|}{$H_2 ~[10^{-1} \ \textrm{GeV}^4]$} & \multicolumn{3}{c|}{$H_3~ [ 10^{-1} \ \textrm{GeV}^6]$} \\ \hline
\multirow{2}{*}{$~\xi_{SM}~$} & $42.959$&$+0.031_{\textrm{pow}}$&$+0.491_{\alpha_s} $ & $2.236$&$+7.089_{\textrm{pow}}$&$+2.043_{\alpha_s} $ & $-0.213$&$+47.209_{\textrm{pow}}$&$-3.632_{\alpha_s} $ \\ 
&& $+0.418_{\frac{\alpha_s}{m_b^2}} $ & $-0.113_{\alpha_s^2}$ & & $-1.146_{\frac{\alpha_s}{m_b^2}} $ & $+1.066_{\alpha_s^2}$ & & $-4.032_{\frac{\alpha_s}{m_b^2}} $ & $-3.975_{\alpha_s^2}$ \\ \hline
$~\xi_{RR}~$ & $-0.213$&$-0.085_{\textrm{pow}}$&$-0.052_{\alpha_s}$ & $0.146$&$-1.072_{\textrm{pow}}$&$-0.054_{\alpha_s}$ & $0.059$&$-0.499_{\textrm{pow}}$&$-0.491_{\alpha_s}$ \\ \hline
$~\xi_{LR}~$ & $2.135$&$-2.023_{\textrm{pow}}$&$+0.019_{\alpha_s}$ & $-0.365$&$+5.087_{\textrm{pow}}$&$+0.011_{\alpha_s}$ & $-0.405$&$+8.232_{\textrm{pow}}$&$-0.395_{\alpha_s}$ \\ \hline
$~\xi_{LR^2}~$ & $1.434$&$-1.477_{\textrm{pow}}$&$-0.011_{\alpha_s}$ & $-0.701$&$+4.301_{\textrm{pow}}$&$+0.003_{\alpha_s}$ & $-0.038$&$+2.071_{\textrm{pow}}$&$-0.266_{\alpha_s}$ \\ \hline
$~\xi_{TT}~$ & $7.917$&$+6.888_{\textrm{pow}}$& & $0.492$&$+12.225_{\textrm{pow}}$& & $-1.625$&$+37.872_{\textrm{pow}}$& \\ \hline
$~\xi_{SS}~$ & $-2.271$&$-2.994_{\textrm{pow}}$& & $0.267$&$-2.385_{\textrm{pow}}$& & $0.464$&$-9.426_{\textrm{pow}}$& \\ \hline
$~\xi_{PP}~$ & $-0.136$&$+0.021_{\textrm{pow}}$& & $-0.098$&$-0.793_{\textrm{pow}}$& & $0.059$&$-1.135_{\textrm{pow}}$& \\ \hline
$~\xi_{ST}~$ & $-0.427$&$+0.542_{\textrm{pow}}$& & $0.292$&$-2.504_{\textrm{pow}}$& & $0.117$&$-1.537_{\textrm{pow}}$& \\ \hline
$~\xi_{PT}~$ & $0.427$&$-0.258_{\textrm{pow}}$& & $-0.292$&$+1.367_{\textrm{pow}}$& & $-0.117$&$+0.863_{\textrm{pow}}$& \\ \hline
    \end{tabular}
    \end{adjustbox}
    \vspace{-0.3cm}
    \caption{Hadronic mass moments with ${E_\ell}_{\rm cut}= 1~\textrm{GeV}$.}
    \label{tab:MX_moments}
\end{table}

\begin{table}[H]
    \centering
    \begin{adjustbox}{max width=\textwidth}
    \begin{tabular}{|c|rll|rll|rll|}
    \hline
 & \multicolumn{3}{c|}{$Q_1 ~[\textrm{GeV}^2]$} & \multicolumn{3}{c|}{$Q_2 ~[\textrm{GeV}^4]$} & \multicolumn{3}{c|}{$Q_3~[\textrm{GeV}^6]$} \\ \hline
\multirow{2}{*}{$~\xi_{SM}~$} & $7.077$&$-0.425_{\textrm{pow}}$&$+0.012_{\alpha_s} $ & $4.293$&$-1.639_{\textrm{pow}}$&$+0.060_{\alpha_s} $ & $3.795$&$-4.475_{\textrm{pow}}$&$+0.477_{\alpha_s} $ \\ 
&$-0.028_{\frac{\alpha_s}{m_b^2}} $ & $-0.024_{\frac{\alpha_s}{m_b^3}} $ & $+0.041_{\alpha_s^2}$ & $-0.050_{\frac{\alpha_s}{m_b^2}} $ & $-0.041_{\frac{\alpha_s}{m_b^3}} $ & $+0.168_{\alpha_s^2}$ & $+0.040_{\frac{\alpha_s}{m_b^2}} $ & $+0.177_{\frac{\alpha_s}{m_b^3}} $ & $+0.578_{\alpha_s^2}$ \\ \hline
$~\xi_{LR}~$ & $-0.693$&$+0.101_{\textrm{pow}}$&$+0.006_{\alpha_s}$ & $-0.769$&$+0.233_{\textrm{pow}}$&$+0.092_{\alpha_s}$ & $2.117$&$-0.768_{\textrm{pow}}$&$+0.165_{\alpha_s}$ \\ \hline
$~\xi_{LR^2}~$ & $-0.681$&$+0.129_{\textrm{pow}}$&$+0.008_{\alpha_s}$ & $-1.236$&$+0.402_{\textrm{pow}}$&$+0.101_{\alpha_s}$ & $0.481$&$-0.129_{\textrm{pow}}$&$+0.360_{\alpha_s}$ \\ \hline
$~\xi_{TT}~$ & $-2.179$&$+0.484_{\textrm{pow}}$& & $-1.598$&$+1.228_{\textrm{pow}}$& & $8.239$&$-1.608_{\textrm{pow}}$& \\ \hline
$~\xi_{SS}~$ & $0.619$&$+0.677_{\textrm{pow}}$& & $0.584$&$+2.189_{\textrm{pow}}$& & $-2.088$&$+4.671_{\textrm{pow}}$& \\ \hline
$~\xi_{PP}~$ & $-0.074$&$+0.004_{\textrm{pow}}$& & $-0.185$&$+0.029_{\textrm{pow}}$& & $0.028$&$-0.018_{\textrm{pow}}$& \\ \hline
    \end{tabular}
    \end{adjustbox}
    \vspace{-0.3cm}
    \caption{$q^2$ moments with $q^2_{\rm cut}= 4~\textrm{GeV}^2$. We have omitted the lines for $\xi_{RR}=\xi_{ST}=\xi_{PT}=0$.}
    \label{tab:q2_moments}
\end{table}

\begin{table}[H]
    \centering
    \begin{tabular}{|l|rlll|}
    \hline 
    &\multicolumn{4}{c|}{$10 \times R^{*}$} \\ \hline
$~\xi_{SM}~$ & $8.175$ & $-0.125_{\textrm{pow}}$ & $-0.001_{\alpha_s} $ & $-0.016_{\alpha_s/m_b^2} -0.004_{\alpha_s^2}$ \\ \hline 
$~\xi_{RR}~$ & $-1.039$&$+0.020_{\textrm{pow}}$&$+0.003_{\alpha_s}$ & \\ \hline
$~\xi_{LR}~$ & $-0.174$&$+0.038_{\textrm{pow}}$&$-0.033_{\alpha_s}$ & \\ \hline
$~\xi_{LR^2}~$ & $-0.113$&$+0.033_{\textrm{pow}}$&$-0.019_{\alpha_s}$ & \\ \hline
$~\xi_{TT}~$ & $-8.316$&$+0.024_{\textrm{pow}}$&& \\ \hline
$~\xi_{SS}~$ & $0.087$&$+0.150_{\textrm{pow}}$ && \\ \hline
$~\xi_{PP}~$ & $-0.087$&$-0.002_{\textrm{pow}}$&& \\ \hline
$~\xi_{ST}~$ & $-2.079$&$-0.081_{\textrm{pow}}$&& \\ \hline
$~\xi_{PT}~$ & $2.079$&$+0.063_{\textrm{pow}}$&& \\ \hline
    \end{tabular}
    \caption{$R^*$ with ${E_\ell}_{\rm cut} = 1~\text{GeV}$.}
    \label{tab:Rstar}
\end{table}

\begin{table}[H]
    \centering
    \begin{tabular}{|l|rl|}
    \hline 
    &\multicolumn{2}{c|}{$10^2 \times A_{FB}$} \\ \hline
$~\xi_{SM}~$ & $18.44$&$-2.08_{\textrm{pow}}$ \\ \hline 
$~\xi_{RR}~$ & $-36.89$&$+4.16_{\textrm{pow}}$ \\ \hline
$~\xi_{LR}$ & $12.00$&$-2.29_{\textrm{pow}}$ \\ \hline
$~\xi_{LR^2}~$ & $7.80$&$-2.10_{\textrm{pow}}$ \\ \hline
$~\xi_{TT}~$ & $~-221.34$&$+31.51_{\textrm{pow}}~$ \\ \hline
$~\xi_{SS}~$ & $-15.22$&$-0.02_{\textrm{pow}}$ \\ \hline
$~\xi_{PP}~$ & $-3.22$&$+0.47_{\textrm{pow}}$ \\ \hline
$~\xi_{ST}~$ & $-73.78$&$-1.22_{\textrm{pow}}$ \\ \hline
$~\xi_{PT}~$ & $73.78$&$-2.88_{\textrm{pow}}$ \\ \hline
 \end{tabular}
    \caption{$A_{FB}$ with lower cut $q^2_{\rm cut} = 4~\text{GeV}$.}
    \label{tab:AFBtab}
\end{table}

\begin{table}[H]
    \centering
    \begin{adjustbox}{max width=\textwidth}
    \begin{tabular}{|c|rlll|}
    \hline 
    &\multicolumn{4}{c|}{$10 \times \Gamma/\Gamma_0$} \\ \hline
$~\zeta_{SM}~$ & $6.580$ & $-0.461_{\textrm{pow}}$ & $-0.560_{\alpha_s} $ & $-0.009_{\alpha_s/m_b^2} -0.034_{\alpha_s/m_b^3} -0.105_{\alpha_s^2}+0.003_{\alpha_s^3}+0.058_{\alpha_{\rm em}}$ \\ \hline 
$~\zeta_{RR}~$ & $6.580$&$-0.461_{\textrm{pow}}$&$-0.560_{\alpha_s}$ & \\ \hline
$~\zeta_{LR}~$ & $-4.280$&$+0.634_{\textrm{pow}}$&$+0.464_{\alpha_s}$ & \\ \hline
$~\zeta_{TT}~$ & $78.964$&$-7.865_{\textrm{pow}}$&& \\ \hline
$~\zeta_{SS}~$ & $5.430$&$+0.240_{\textrm{pow}}$&& \\ \hline
$~\zeta_{PP}~$ & $1.150$&$-0.118_{\textrm{pow}}$&& \\ \hline
    \end{tabular}
    \end{adjustbox}
    \vspace{-0.3cm}
    \caption{Total decay width $\Gamma$ according to Eq.~\eqref{eq:gammatildes}.}
    \label{tab:Gammatot}
\end{table}

\section{Additional plots}
\label{sec:appplots}
In this section we collect the 1D profile $\Delta \chi^2$ plots for the NP parameters and $|\tilde{V}_{cb}|$ in the single-mediator scenarios outlined in Section~\ref{sec:globfit}, and in the full NP case.
The plots are obtained by minimizing the $\chi^2$ for some values of the abscissa (black points) and by interpolating the results (solid lines).
Central values, with $\Delta \chi^2 = 1$ intervals, are shown on the plots.
The minimum $\chi^2$ are reported in~\eqref{eq:chimins}.
\begin{figure}[H]
    \centering
    \includegraphics[width=0.9\linewidth]{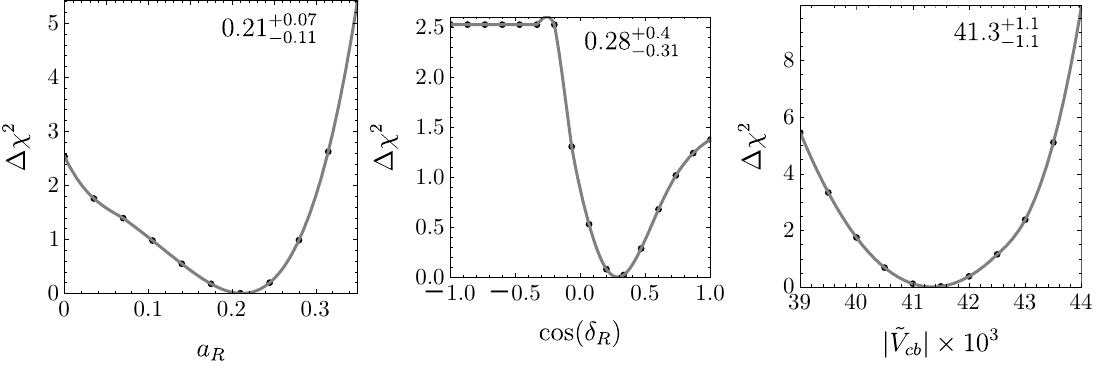}
    \vspace{-0.6cm}
    \caption{Fit results for the NP parameters and $|\tilde{V}_{cb}|$ in scenario I.}
    \label{fig:1DI}
\end{figure}
\begin{figure}[H]
    \centering
    \includegraphics[width=\linewidth]{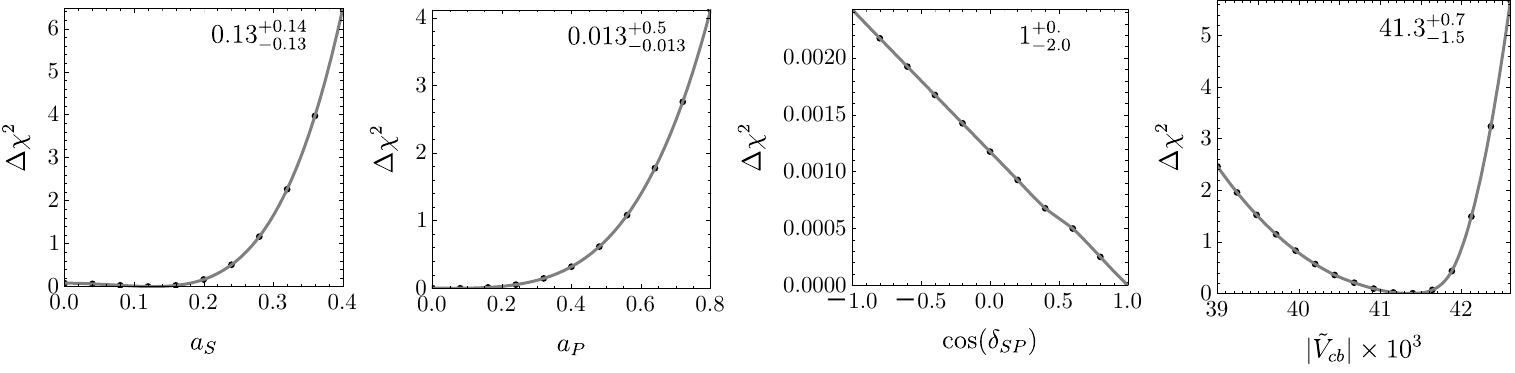}
    \vspace{-0.9cm}
    \caption{Fit results for the NP parameters and $|\tilde{V}_{cb}|$ in scenario II.}
    \label{fig:1DII}
\end{figure}
\begin{figure}[H]
    \centering
    \includegraphics[width=0.65\linewidth]{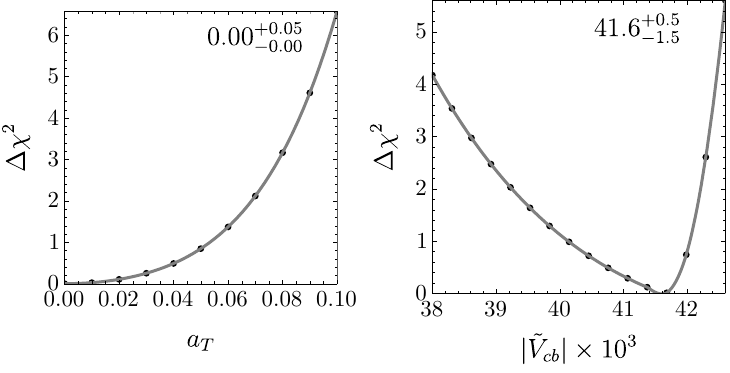}
    \vspace{-0.5cm}
    \caption{Fit results for the NP parameters and $|\tilde{V}_{cb}|$ in scenario III.}
    \label{fig:1DIII}
\end{figure}
\begin{figure}[H]
    \centering
    \includegraphics[width=0.65\linewidth]{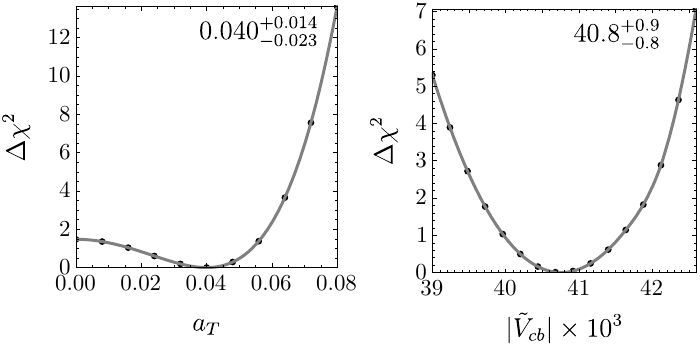}
    \vspace{-0.5cm}
    \caption{Fit results for the NP parameters and $|\tilde{V}_{cb}|$ in scenario IV.}
    \label{fig:1DIV}
\end{figure}
\begin{figure}[H]
    \centering
    \includegraphics[width=0.65\linewidth]{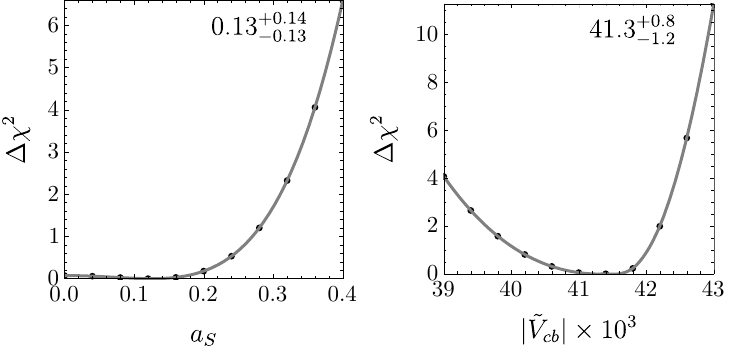}
    \vspace{-0.5cm}
    \caption{Fit results for the NP parameters and $|\tilde{V}_{cb}|$ in scenario V.}
    \label{fig:1DV}
\end{figure}
\begin{figure}[H]
    \centering
    \includegraphics[width=\linewidth]{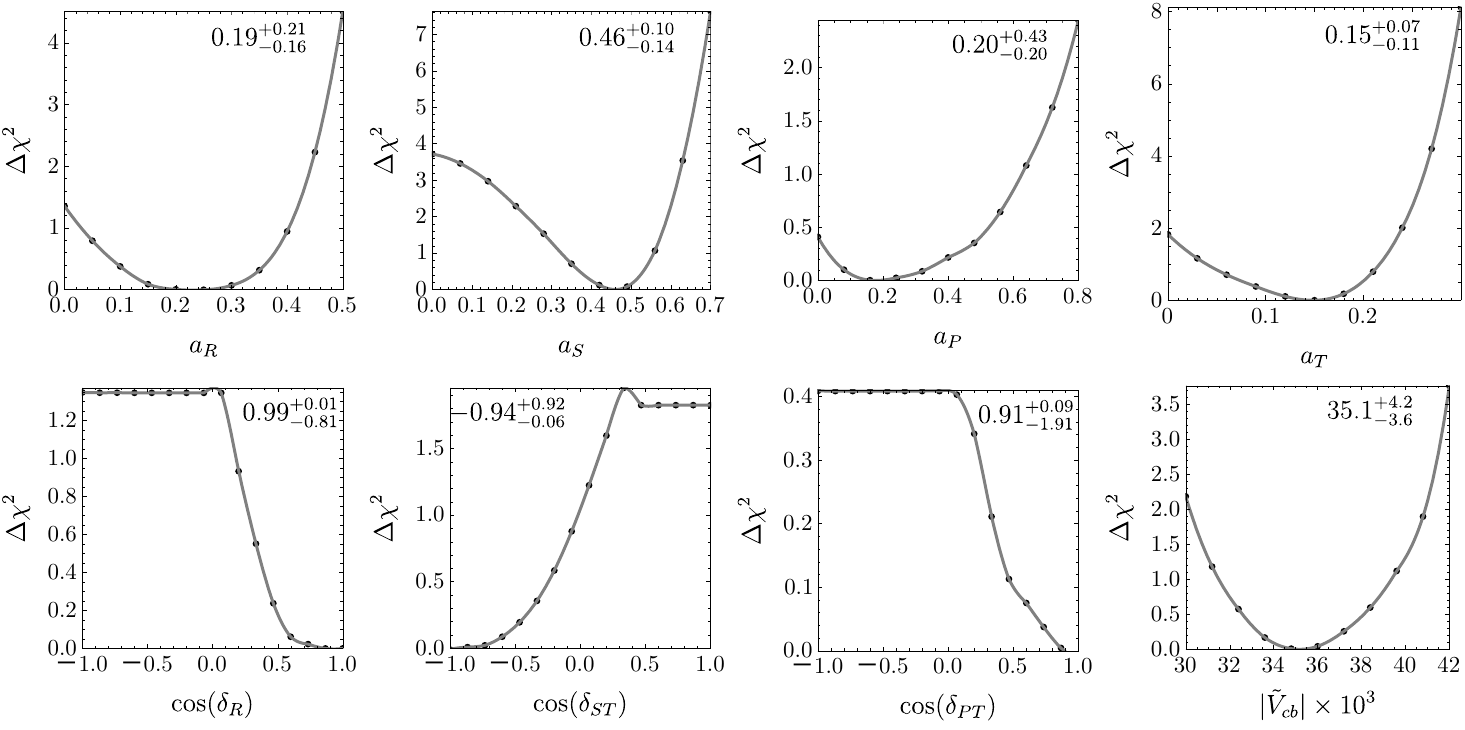}
    \vspace{-1cm}
    \caption{Fit results for the NP parameters and $|\tilde{V}_{cb}|$ in the full NP case.}
    \label{fig:1DfullNP}
\end{figure}

\bibliographystyle{JHEP.bst}
\bibliography{bibiliography.bib}

\end{document}